\documentclass[11pt,letter]{article}
\newtheorem{theorem}{Theorem}[section]
\newtheorem{definition}{Definition}[section]

\newtheorem{lemma}{Lemma}[section]
\newtheorem{proposition}{Proposition}[section]

\usepackage{amsmath}
\usepackage{amssymb}

\begin{document}

\title{Concurrent Knowledge  Extraction in the Public-Key Model\thanks{This work was being under a journal submission since August 2008, where the journal policy  prohibits opening  the work until getting the review decisions. Unfortunately, we did not get any review response up to now since the submission. We decide to withdraw the journal submission, and make it open here. The version opened here remains  intact with the journal submission made in August 2008.
 This
 work 
  was supported in part
    by a grant
from the Research Grants Council of the Hong Kong Special
Administrative Region, China (No.  CityU 122105),  CityU
Research Grant (No. 9380039) and  a grant from the 
Basic Research Development (973) Program of China 
(No. 2007CB807901).
 The third author is also supported by 
 NSFC (No. 60703091), the Pujiang
Program of Shanghai and a grant from MSRA.}}


\author{Andrew C. C. Yao\footnote{Institute for Theoretical Computer Science (ITCS), Tsinghua University, Beijing, China.
\texttt{andrewcyao@tsinghua.eud.cn} Works partially done while
 visiting City University of Hong Kong.}
\and Moti Yung\footnote{Google Inc.  and  Columbia University, New
York, NY, USA. \quad \texttt{moti@cs.columbia.edu}} \and Yunlei
Zhao\footnote{Corresponding author. Software School, Fudan
University, Shanghai 200433,
 China.    \texttt{ylzhao@fudan.edu.cn} Works partially done while visiting Tsinghua university and
City University of Hong Kong. }}
\date{}

\date{}          
\maketitle

\begin{abstract}

Knowledge extraction is a fundamental notion, modelling machine
possession of values (witnesses) in a computational complexity
sense. The notion provides an essential tool for cryptographic
protocol design and analysis, enabling one to argue about the
internal state of protocol players without ever looking at this
supposedly secret state.
%
%
However, when transactions are concurrent (e.g., over the
Internet) with players possessing public-keys (as is common in
cryptography), assuring that entities ``know'' what they claim to
know, where adversaries may be well coordinated across different
transactions, turns out to be much more subtle and in need of
re-examination.  Here, we investigate how to formally treat
knowledge possession by parties (with registered public-keys)
interacting over the Internet.  Stated more technically, we look
into the relative power of the notion of ``concurrent
knowledge-extraction'' (CKE) in the concurrent zero-knowledge
(CZK) bare public-key (BPK) model.

We show the potential vulnerability of  man-in-the-middle (MIM)
attacks turn out to be a real security threat to existing natural
protocols  running concurrently in the 
 public-key model, which
motivates us to introduce and formalize the notion of CKE. 
%
 Then,   both generic (based on standard polynomial assumptions) and efficient (employing complexity
 leveraging in a novel way) implementations for $\mathcal{NP}$ are presented for
constant-round (in particular, round-optimal) concurrently
knowledge-extractable concurrent zero-knowledge (CZK-CKE) arguments
 in the BPK model. The efficient implementation can be further high practically
 instantiated  for specific number-theoretic language. Along the
 way, we discuss and clarify the various subtleties surrounding the  security
 formulation and analysis, which provides 
  insights into
  the complex CZK-CKE setting.

\end{abstract}


\section{Introduction}       

Zero-knowledge (ZK) protocols allow a prover to assure a verifier
of validity of theorems without giving away any additional
knowledge (i.e., computational advantage) beyond validity.  This
notion was introduced by Goldwasser, Micali and Rackoff
\cite{GMR85} and its generality was demonstrated by Goldreich,
Micali and Wigderson \cite{GMW91}. Since its introduction ZK has
found numerous useful applications, and by now has been playing a
central role for modern cryptography  (particularly  in
cryptographic protocol design \cite{Y86,GMW87}).

Traditional notion of ZK considers the security in a stand-alone
(or sequential) execution of the protocol. Motivated by the use of
such protocols in an asynchronous network like the Internet, where
many protocols  run simultaneously,
studying security properties of ZK protocols in such  concurrent
settings has attracted much research efforts in recent years,
starting by Dwork, Naor and Sahai \cite{DNS98}.  Informally, a ZK
protocol is called concurrent zero-knowledge (CZK) if concurrent
instances are all expected polynomial-time simulatable,
namely, when a possibly malicious verifier concurrently interacts
with a polynomial number of honest prover instances and schedules
message exchanges as it wishes.

The concept of ``proof of knowledge'' (POK), informally discussed
in \cite{GMR85},  was then formally treated
(see \cite{FFS88,BG92,G01,BG06}). 
 POK systems, especially zero-knowledge POK (ZKPOK) systems, play a
fundamental role in the design of cryptographic schemes and
protocols, enabling a formal complexity theoretic treatment of
what does it mean for a machine to ``know'' something.  Roughly
speaking, a ``proof of knowledge'' means that a possibly malicious
prover can convince the verifier that an $\mathcal{NP}$ statement
is true if and only if it, in fact, ``knows'' (i.e., possesses) a
witness to the statement (rather than merely conveying ``proof of
language membership,'' i.e., the fact that a corresponding witness
exists).

With the advancement of cryptographic models where parties
initially publish public-keys (particularly for achieving
round-efficient concurrently secure protocols \cite{CKPR02}),
knowledge extraction becomes more subtle (due to possible
dependency on published keys), and needs re-examination.  Here, we
investigate the relative power of the notion of ``concurrent
knowledge-extraction'' in the concurrent zero-knowledge  bare
public-key model. Namely, we investigate how to formally treat
knowledge possessions for parties (which own public-keys)
interacting over the Internet.


The bare public-key (BPK) model, originally introduced by Canetti,
Goldreich, Goldwasser and Micali   \cite{CGGM00},
%
%
 is a natural and relatively weak
 cryptographic model.  A protocol in this model
 simply assumes that all verifiers have each deposited a public key in
 a public file
%
%
before (or while)
 user interactions  take place.
%
%
 No assumption is made on whether the public-keys
deposited are unique or valid (i.e., public keys can even be
``nonsensical,"  where no corresponding secret-keys exist or are
known). That is, no trusted third party is assumed, the underlying
communication network is assumed to be adversarially asynchronous
(i.e., arbitrary message delays), and
preprocessing is reduced to minimally non-interactively posting
public-keys in a public file
 (dynamic posting is allowed assuming a reasonable amount of time
between key posting and key usage \cite{CGGM00}).
%
%
%
 In many cryptographic settings, availability of a public key
infrastructure (PKI) is assumed or required,  and in these
settings the BPK model is, both, natural and attractive (note that
the BPK model is, in fact, a weaker version of PKI where in the
later added key certification is assumed).  It was pointed out by
Micali and Reyzin \cite{MR01a} that the BPK model is, in fact,
applicable to interactive systems in general.

Verifier security (i.e., soundness) in the BPK model (against
malicious provers) turned out to be more involved
 than anticipated, as was demonstrated by Micali and Reyzin \cite{MR01a}
who showed that under standard intractability assumptions there
are four distinct meaningful notions of soundness, i.e., from
weaker to stronger: one-time, sequential, concurrent and
resettable soundness. Here, we focus on concurrent soundness,
which, roughly speaking, means that a possibly malicious
probabilistic polynomial-time (PPT) prover $P^*$ cannot convince
the honest verifier $V$ of a \emph{false} statement even when
$P^*$ is allowed multiple interleaving interactions with $V$ in
the public-key model. They also showed that any black-box ZK
protocol with concurrent soundness in the BPK model (for
non-trivial languages outside $\mathcal{BPP}$) must run at least
four rounds \cite{MR01a}. It was also shown in \cite{BGGL01,MR01a}
that \emph{black-box} ZK arguments with resettable soundness only
exist for trivial (i.e, $\mathcal{BPP}$) languages (whether in the
BPK model or not).

Due to the above, it was    implied  that concurrent soundness
might be the best verifier security one can hope for in the case
of black-box ZK arguments in the BPK model. In this work, we show
that this intuition is not entirely correct, at least not in the
POK setting
 where provers are polynomial time.
Specifically, concurrent soundness only guarantees that
concurrently interleaved interactions cannot help a malicious
prover validate a \emph{false} statement in the public-key  model.
However, it does \emph{not} prevent a malicious prover from
validating a \emph{true} statement {\em but} without knowing any
witness for the statement being proved.
One reason that this potential vulnerability is not merely a
theoretical concern is that:  all concurrent ZK protocols in the
BPK model involve a sub-protocol in which the verifier proves to
the prover the knowledge of the secret-key corresponding to its
registered public-key; Further, this type of proofs are also quite
common in practical cryptographic protocols in the public-key
model.
%
%
  A malicious prover, in turn, can potentially exploit these proofs by
the verifier in other sessions, without possessing a witness to
these sessions' statements. 
%
%
%
We show  concrete instances of this  vulnerability.
%
%
This issue, therefore, motivates the need for careful definitions
and
 for achieving concurrent verifier security for concurrent ZK POK in
 the BPK model, so that \emph{provably} one can remedy the above security
 vulnerability.

%

\subsection{Our contributions}

We start by investigating the subtleties of concurrent verifier
 security in the public-key model in the case of proof of
 knowledge. Specifically, we show  concurrent interleaving and
 malleating attacks against some existing natural protocols running concurrently in the
 BPK model, which shows that concurrent soundness and normal arguments of knowledge (and also traditional
 concurrent non-malleability)  do not
 guarantee concurrent verifier security in the BPK model. 

Then, we formulate concurrent verifier security that remedies the
vulnerability as demonstrated by the concrete  attacks which are of
the man-in-the-middle nature.   
The security notion defined is named \texttt{concurrent
knowledge-extraction (CKE)} in the public-key model, which
essentially means that for statements   whose validations are
successfully conveyed by a possibly malicious prover to an honest
verifier (with registered public-key) by concurrent interactions,
the prover must ``know" the corresponding witnesses. 
We then  present  both generic (based on standard polynomial
assumptions) and efficient (employing complexity
 leveraging in a novel way) implementations of constant-round (in particular, round-optimal)
 CZK-CKE arguments for $\mathcal{NP}$ in the BPK model. 
 The efficient implementation can be further high practically
 instantiated  for specific number-theoretic language.
The techniques developed in this work for achieving CZK and CKE
simultaneously could  be  of independent interests. Specifically,
although some non-malleable building tools  seem 
   to be intrinsically required  for achieving CZK-CKE in the BPK model, our
solution does not employ any non-malleable tools.
 Along
the
 way, we discuss and clarify the  various subtleties surrounding the  security formulation
 and analysis, which provides insights into 
  the complex CZK-CKE setting. 

As knowledge-extraction and zero-knowledge (and also the
public-key model)
  are fundamental to cryptography,  we suggest
  that the clarifications and formulation of CKE in the public-key
  model,  the (both generic and efficient) CZK-CKE constructions and techniques developed  in this work,
  along with the  discussions and clarifications of the various subtleties surrounding the security formulation and
  analysis,   are 
  fundamental and can  serve as a basis  to
 formulate and achieve 
 more complex  cryptographic protocols
 in the public-key model. In particular, the CZK-CKE protocols are
 themselves  the
 concurrent version, in the public-key model,  of the highly useful and fundamental  zero-knowledge
  arguments of
 knowledge.

\subsection{Related works} \label{relatedwork}

Let us review some recent results and developments; we have been
 involved in numerous recent works which we review together with
 related works.  While the list of related works and related issues
 is  quite lengthy, the bottom line is that the notion defined and
 achieved herein is unique and independent of various related issues
 and works, and it captures knowledge extraction as a basic issue in
 concurrent executions in public key models.

Concurrent ZK (actually, resettable ZK that is stronger than CZK)
arguments for $\mathcal{NP}$ with a \emph{provable
sub-exponential-time} CKE property in the BPK model were first
achieved in  \cite{YZ05}, which make sense only for
sub-exponentially hard languages.
 Standard \emph{polynomial-time} CKE for
concurrent ZK arguments in the BPK model  were left over there as
an open problem, which we answer here. We note that the techniques
used in \cite{YZ05} do not render CZK with \emph{polynomial-time}
concurrent knowledge-extraction, and the subtle  issues of
knowledge-extraction independence were  not realized and
formalized
there. 

Two constructions for concurrent ZK arguments with sequential
soundness in the BPK model under standard assumptions were
proposed in the incomplete work of \cite{Z03} (the early version
since January 2004). But, the security proof of concurrent
soundness turned out to be flawed, as observed independently in
\cite{DV05,YZ06}. One construction was fixed to be concurrently
sound in \cite{DV05} by introducing some new techniques, and
recently another construction was fixed to be concurrently sound
in \cite{DL06} following the spirit of \cite{DV05}.  Given these
works, the current work (with its preliminary version appeared in
\cite{YYZ07eccc}) further shows that the concurrently sound CZK
arguments of \cite{DV05,DL06} do not capture CKE and are not
concurrently knowledge-extractable when it comes to proofs of
knowledge.

Recently  in another separate  work \cite{YYZ07}, which deals with
concurrent non-malleability (CNM)  in the BPK model, we further
clarify that the formulations of
 concurrent non-malleability (CNM) in existing works \cite{OPV06,DDL06}
 do not capture CKE in the public-key model. (Note that the preliminary version of this work, appeared in
 August-2006 update of the incomplete work of \cite{Z03}, is independent of \cite{OPV06,DDL06}.)
 It is also demonstrated  there  that  
 the CNMZK protocol of \cite{DDL06} is not concurrently
 knowledge-extractable (in the sense that concrete attacks exist).
   The  line of CNM explorations in the BPK model is outside of
  the scope of the current
work. 

 In general, the issue of concurrent composition of proof
 of knowledge (POK) could be traced back to the  works \cite{DDN00,GKr96}. 

\subsection{Organization}
We recall basic notions and   tools  in Section \ref{basic}. In
Section \ref{APPBPK}, we describe (an augmented version) of the
BPK model with adaptive language selections based on public-keys.
In Section \ref{appattack}, we present the motivation, by concrete
attacks on naturally existing protocol,  for concurrent
knowledge-extractability in the public-key model. 
In Section \ref{CKEformulation}, we formulate CKE in the BPK
model, and make   clarifications and justification of  the CKE
formulation.
In Section \ref{genericCKE}, we present the generic implementation
of constant-round CZK-CKE arguments for $\mathcal{NP}$ in the BPK
model under standard hardness assumptions.  In Section
\ref{efficient}, we present the efficient and practical
implementations of constant-round CZK-CKE arguments for
$\mathcal{NP}$ in the BPK model with the usage of complexity
leveraging in a minimal and  novel way,  and discuss and clarify
in depth  the various subtleties. 


\section{Preliminaries}\label{basic}

We use standard notations and conventions below for writing
probabilistic algorithms, experiments and interactive protocols.
If \emph{A} is
 a probabilistic algorithm, then $A(x_1, x_2, \cdots; r)$ is the result of running \emph{A} on inputs
 $x_1, x_2, \cdots$ and coins $r$. We let $y\leftarrow A(x_1, x_2, \cdots)$ denote the experiment of picking $r$ at
 random and letting $y$ be $A(x_1, x_2, \cdots; r)$. If $S$ is a finite set then $x\leftarrow S$ is the operation of
 picking an element uniformly from $S$. If $\alpha$ is neither an algorithm nor a set then $x\leftarrow \alpha$ is a
 simple assignment statement.  By $[R_1; \cdots; R_n: v]$ we
 denote the set of values of $v$ that a random variable can
 assume, due to the distribution determined by the sequence of
 random processes $R_1, R_2, \cdots, R_n$. By $\Pr[R_1; \cdots; R_n: E]$ we denote
 the probability of event $E$, after the ordered execution of
 random processes $R_1, \cdots, R_n$.

 Let $\langle P, V\rangle$ be a
 probabilistic
 interactive protocol, then the notation $(y_1, y_2)\leftarrow
 \langle P(x_1), V(x_2)\rangle (x)$ denotes the random process of
 running interactive protocol $\langle P, V \rangle$ on common input $x$, where $P$
 has private input $x_1$, $V$ has private input $x_2$, $y_1$ is
 $P$'s output and $y_2$ is $V$'s output. We assume w.l.o.g. that the
 output of both parties $P$ and $V$ at the end of an execution of
 the protocol $\langle P, V \rangle$ contains a transcript of the
 communication exchanged between $P$ and $V$ during such
 execution.

The security  of cryptographic primitives and tools presented in
this section is defined with respect to uniform polynomial-time or
sub-exponential-time algorithms (equivalently, polynomial-size or
sub-exponential-size circuits). When it comes to non-uniform
security, we refer to non-uniform polynomial-time or
sub-exponential-time algorithms (equivalently, families of
circuits of polynomial or sub-exponential size).

\begin{definition}[one-way
function]\label{OWF} A function $f: \{0, 1\}^*\longrightarrow \{0,
1\}^*$ is called a  one-way function (OWF) if the following
conditions hold:
\begin{enumerate}
\item Easy to compute: There exists a (deterministic)
polynomial-time algorithm $A$ such that on input $x$ algorithm $A$
outputs $f(x)$ (i.e., $A(x)=f(x)$).

\item Hard to invert: For every probabilistic polynomial-time PPT
algorithm $A^{\prime}$, every positive polynomial $p(\cdot)$, and
all sufficiently large $n$'s, it holds $\Pr[A^{\prime}(f(U_n),
1^n)\in f^{-1}(f(U_n))]<\frac{1}{p(n)}$, where $U_n$ denotes a
random variable uniformly distributed over $\{0, 1\}^n$. A OWF $f$
is called \emph{sub-exponentially strong} if for some constant
$c$, $0<c<1$, for every sufficiently large $n$, and every circuit
$C$ of size at most $2^{n^c}$, $\Pr[C(f(U_n), 1^n)\in
f^{-1}(f(U_n))]<2^{-n^c}$.

\end{enumerate}

\end{definition}

\begin{definition}[(public-coin) interactive argument/proof system]
A pair of  interactive  machines, $\langle P,\,V \rangle$, is
called an interactive argument  system for a language $L$ if both
are probabilistic polynomial-time (PPT) machines and  the
following conditions hold:
\begin{itemize}
\item Completeness. For every $x\in L$, there exists a string $w$
such that for every string $z$,  \\ $\Pr[\langle P(w),\,V(z)
\rangle(x)=1]=1$. \item Soundness. For every polynomial-time
interactive machine $P^*$, and for all sufficiently large $n$'s
and every $x\notin L$ of length $n$ and every $w$ and $z$,
$\Pr[\langle P^*(w),\,V(z) \rangle(x)=1]$ is negligible in $n$.
\end{itemize}
An interactive protocol is called a \emph{proof} for $L$, if the
soundness condition holds against any (even power-unbounded) $P^*$
(rather than only PPT $P^*$). An interactive  system is called  a
public-coin  system if at each round the prescribed verifier can
only toss coins and send their outcome  to the prover.
\end{definition}

Commitment schemes enable a party, called the \emph{sender}, to
bind itself to a value in the initial \emph{commitment} stage,
while decurving it from the \emph{receiver} (this property is
called \emph{hiding}). Furthermore, when the commitment is opened
in a later \emph{decommitment} stage, it is guaranteed that the
``opening'' can yield only the single value determined in the
commitment phase (this property is called \emph{binding}).
Commitment schemes come in two different flavors:
statistically-binding computationally-hiding and
statistically-hiding computationally-binding.

\begin{definition} [statistically/perfectly binding bit commitment scheme]
A  pair of PPT interactive machines,
 $\langle P, V\rangle$, is called a perfectly binding bit commitment scheme, if it satisfies  the following:

\begin{description}
\item [Completeness.] For any security parameter $n$,  and any bit
$b\in \{0, 1\}$, it holds that \\ $\Pr[(\alpha, \beta)\leftarrow
\langle P(b), V\rangle (1^n); (t, (t, v))\leftarrow \langle
P(\alpha), V(\beta)\rangle(1^n): v=b]=1$.

\item [Computationally hiding.]  For all sufficiently large $n$'s,
any PPT adversary $V^*$, the following two probability
distributions are  computationally indistinguishable: $[(\alpha,
\beta)\leftarrow \langle P(0), V^*\rangle (1^n): \beta]$  and
$[(\alpha^{\prime}, \beta^{\prime})\leftarrow \langle P(1),
V^*\rangle (1^n):
\beta^{\prime}]$. 

\item [Perfectly Binding.] For all sufficiently large $n$'s, and
\emph{any}  adversary $P^*$, the following probability is
negligible (or equals 0 for perfectly-binding commitments):
$\Pr[(\alpha, \beta)\leftarrow \langle P^*, V\rangle (1^n); (t,
(t, v))\leftarrow \langle P^*(\alpha), V(\beta)\rangle(1^n);
(t^{\prime}, (t^{\prime}, v^{\prime}))\leftarrow \langle
P^*(\alpha), V(\beta)\rangle(1^n): v, v^{\prime} \in \{0, 1\}
\bigwedge v\neq v^{\prime}]$.

That is, no (\emph{even computational power unbounded}) adversary
$P^*$ can decommit the same transcript of the commitment stage
both to 0 and 1.

\end{description}

\end{definition}

Below, we recall some classic perfectly-binding commitment
schemes.

 One-round
perfectly-binding (computationally-hiding) commitments can be
  based on any one-way permutation OWP \cite{B82,GMW91}.
Loosely speaking, given a OWP $f$ with a hard-core predict $b$
(cf. \cite{G01}), on a security parameter $n$ one commits  a bit
$\sigma$ by uniformly selecting $x\in \{0, 1\}^n$ and sending
$(f(x), b(x)\oplus \sigma)$ as a commitment, while keeping $x$ as
the decommitment information.

 For practical perfectly-binding commitment scheme, in this work we use
the DDH-based ElGamal (non-interactive) commitment scheme
\cite{E85}. To commit to a value $v\in Z_q$, the committer
randomly selects $u, r \in Z_q$, computes $h=g^u \mod p$ and sends
$(h, \bar{g}=g^r, \bar{h}=g^vh^r)$ as the commitment. The
decommitment information is $(r, v)$. Upon receiving the
commitment $(h, \bar{g}, \bar{h})$, the receiver checks that $h,
\bar{g}, \bar{h}$ are elements of order $q$ in  $Z^*_p$. It is
easy to see that the commitment scheme is of perfectly-binding.
The computational hiding property is from the DDH assumption on
the subgroup of order $q$ of $Z^*_p$ (for more details, see
\cite{E85}). We also note that in \cite{MP03} Micciancio and
Petrank presented another implementation of DDH-based
perfectly-binding commitment scheme with advanced security
properties. 

Statistically-binding commitments can be  based on any one-way
function (OWF) but run in two rounds \cite{N91,HILL99}. On a
security parameter $n$, let $PRG: \{0, 1\}^n\longrightarrow \{0,
1\}^{3n}$ be a pseudorandom generator, the Naor's OWF-based
two-round public-coin perfectly-binding commitment scheme works as
follows: In the first round, the commitment receiver sends a
random string $R\in \{0, 1\}^{3n}$ to the committer. In the second
round, the committer uniformly selects a string $s\in \{0, 1\}^n$
at first; then to commit  a bit 0 the committer sends $PRG(s)$ as
the commitment; to commit  a bit 1 the committer sends
$PRG(s)\oplus R$ as the commitment. Note that the first-round
message of Naor's commitment scheme can be fixed once and for all
and, in particular, can be posted as a part of public-key in the
public-key model.

\begin{definition} [trapdoor bit commitment  scheme]
A  trapdoor bit commitment   scheme (TC) is a quintuple of
probabilistic polynomial-time (PPT) algorithms TCGen, TCCom,
TCVer, TCKeyVer and TCFake, such that
\begin{description}
\item [Completeness.] For any security parameter $n$,  and any bit
$b\in \{0, 1\}$, it holds that:
\\ $\Pr[(TCPK,
TCSK)\leftarrow\textup{TCGen}(1^n); (c,
d)\leftarrow\textup{TCCom}(1^n,  TCPK, b):
\\  \textup{TCKeyVer}(1^n, TCPK)=\textup{TCVer}(1^n,TCPK, c, b,
d)=1]=1$.

 \item [Computationally Binding.] For all
sufficiently large $n$'s and for any PPT adversary $A$, the
following probability is negligible in $n$:   $\Pr[(TCPK,
TCSK)\leftarrow\textup{TCGen}(1^n); (c, v_1, v_2, d_1,
d_2)\leftarrow\textup{A}(1^n,  TCPK):\\
\textup{TCVer}(1^n,  TCPK, c, v_1, d_1)=\textup{TCVer}(1^n, TCPK,
c, v_2, d_2)=1 \bigwedge v_1, v_2\in \{0, 1\} \bigwedge v_1\neq
v_2 ]$.

\item [Perfectly (or computationally) Hiding.] For all
sufficiently large $n$'s and any   $TCPK$ such that
$\textup{TCKeyVer}(1^n, TCPK)=1$, the following two probability
distributions are
 identical (or computationally indistinguishable):  $[(c_0,
d_0)\leftarrow\textup{TCCom}(1^n,  TCPK, 0): c_0]$ and \\ $[(c_1,
d_1)\leftarrow\textup{TCCom}(1^n,  TCPK, 1): c_1]$.

 \item [Perfect (or Computational) Trapdoorness.] For all
sufficiently large $n$'s and any   $(TCPK, TCSK)\in
\{\textup{TCGen}(1^n)\}$, $\exists v_1\in \{0, 1\}$, $\forall
v_2\in \{0, 1\}$ such that the
following two probability distributions are identical (or computationally indistinguishable):\\
$[(c_1, d_1)\leftarrow\textup{TCCom}(1^n,  TCPK, v_1);
d_2^{\prime}\leftarrow\textup{TCFake}(1^n,  TCPK, TCSK, c_1, v_1,
d_1, v_2): (c_1, d_2^{\prime})]$ and $[(c_2,
d_2)\leftarrow\textup{TCCom}(1^n,  TCPK, v_2): (c_2, d_2)]$.
\end{description}
\end{definition}

\textbf{Feige-Shamir  trapdoor commitments (FSTC) \cite{FS89}.}
 Based on Blum's protocol for DHC, Feige and Shamir developed a
 generic  (computationally-hiding and
 computationally-binding) trapdoor commitment scheme \cite{FS89},
 under either any one-way permutation  or any OWF (depending on the
 underlying perfectly-binding commitment scheme used).  The $TCPK$ of the FSTC scheme  is $(y=f(x), G)$
 (for OWF-based solution,
 $TCPK$ also includes a random string $R$ serving as the first-round message of Naor's OWF-based perfectly-binding commitment
 scheme), where $f$ is a OWF and
 $G$ is a graph that is reduced from $y$ by the Cook-Levin $\mathcal{NP}$-reduction. The corresponding
 trapdoor is $x$ (or equivalently,
 a Hamiltonian cycle in $G$). The following is the description of the Feige-Shamir trapdoor bit commitment
 scheme, on a security parameter $n$.

\begin{description}
 \item [Round-1.] Let $f$ be a OWF, the commitment receiver
 randomly selects an element $x$ of length $n$ in the domain of $f$, computes
 $y=f(x)$, reduces $y$ (by Cook-Levin $\mathcal{NP}$-reduction) to an instance of DHC, a graph $G=(V, E)$ with $q=|V|$ nodes, such
 that finding a Hamiltonian cycle in $G$ is equivalent to finding
 the preimage  of $y$. Finally, it sends $(y, G)$ to the
 committer. We remark that to get OWF-based trapdoor
 commitments, the commitment receiver also sends a random string
 $R$ of length $3n$.
 \item [Round-2.]  The committer first checks the $\mathcal{NP}$-reduction from $y$ to $G$ and aborts if
 $G$ is not reduced from $y$. Otherwise, to commit to $0$, the
 committer selects a random permutation, $\pi$,
of the vertices $V$, and commits (using the underlying
perfectly-binding commitment scheme) the entries of the adjacency
matrix of the resultant permutated graph. That is, it sends an
$q$-by-$q$ matrix of commitments so that the $(\pi(i),
\pi(j))^{th}$ entry is a commitment to 1 if $(i, j)\in E$, and is
a commitment to 0 otherwise; To commit to 1, the committer commits
an adjacency matrix containing a randomly labeled $q$-cycle only.

\item [Decommitment stage.] To decommit to $0$, the committer
sends $\pi$ to the commitment receiver  along with the revealing
of all commitments, and the receiver checks that the revealed
graph is indeed isomorphic to $G$ via $\pi$; To decommit to 1, the
committer only opens the entries of the adjacency matrix that are
corresponding to the randomly labeled cycle, and the receiver
checks that all revealed values are 1 and the corresponding
entries form a simple $q$-cycle.
 \end{description}


\begin{definition}[witness indistinguishability
WI]\label{definitionWI} Let $\langle P,\,V \rangle$ be an
interactive system for a language $L\in \mathcal{NP}$, and let
$R_L$ be the fixed $\mathcal{NP}$ witness relation for $L$. That
is, $x\in L$ if there exists a $w$ such that $(x,\,w)\in R_L$. We
denote by $view_{V^*(z)}^{P(w)}(x)$ a random variable describing
the transcript of all messages exchanged between a (possibly
malicious) PPT verifier $V^*$ and the honest prover $P$ in an
execution of the protocol on common input $x$, when $P$ has
auxiliary input $w$ and $V^*$ has auxiliary input $z$. We say that
$\langle P,\,V \rangle$ is witness indistinguishable for $R_L$ if
for every PPT interactive machine $V^*$, and every two sequences
$W^1=\{w^1_x\}_{x\in L}$ and $W^2=\{w^2_x\}_{x\in L}$ for
sufficiently long $x$, so that $(x,\,w^1_x)\in R_L$ and $(x,\,
w^2_x)\in R_L$, the following two probability distributions are
computationally indistinguishable by any non-uniform
polynomial-time algorithm:
$\{x,\,view_{V^*(z)}^{P(w^1_x)}(x)\}_{x\in L,\, z\in \{0,\,1\}^*}$
and $\{x,\,view_{V^*(z)}^{P(w^2_x)}(x)\}_{x\in L,\, z\in
\{0,\,1\}^*}$. Namely, for every  non-uniform polynomial-time
distinguishing algorithm $D$, every polynomial $p(\cdot)$, all
sufficiently long $x\in L$, and all $z\in \{0, 1\}^*$, it holds
that
$$|\Pr[D(x, z, view_{V^*(z)}^{P(w^1_x)}(x)=1]-\Pr[D(x, z,
view_{V^*(z)}^{P(w^2_x)}(x)=1]|<\frac{1}{p(|x|)}$$

\end{definition}

\begin{definition}[strong witness indistinguishability SWI]
Let $\langle P, V\rangle$ and all other notations be as in
Definition \ref{definitionWI}. We say that $\langle P, V\rangle$
is \emph{strongly witness-indistinguishable for $R_L$} if for
every PPT interactive machine $V^*$ and for every two probability
ensembles $\{X^1_n, Y^1_n, Z^1_n\}_{n\in N}$ and $\{X^2_n, Y^2_n,
Z^2_n\}_{n\in N}$, such that each $\{X^i_n, Y^i_n, Z^i_n\}_{n\in
N}$ ranges over $(R_L\times \{0, 1\}^*)\cap (\{0, 1\}^n \times
\{0, 1\}^* \times \{0, 1\}^*)$, the following holds: If $\{X^1_n,
Z^1_n\}_{n\in N}$ and $\{X^2_n,  Z^2_n\}_{n\in N}$ are
computationally indistinguishable, then so are  $\{\langle
P(Y^1_n), V^*(Z^1_n)\rangle (X^1_n)\}_{n\in N}$ and $\{\langle
P(Y^2_n), V^*(Z^2_n)\rangle (X^2_n)\}_{n\in N}$.

\end{definition}

\textbf{WI vs. SWI:} It is clarified in \cite{G02} that the notion
of SWI actually refers to issues that are fundamentally different
from WI. Specifically, the issue is whether the interaction with
the prover helps $V^*$ to distinguish some auxiliary information
(which is indistinguishable without such an interaction).
Significantly different from WI, SWI does \emph{not} preserve
under concurrent composition. More details about SWI are referred
to \cite{G02}. But, an interesting observation is: the protocol
composing commitments and SWI can be  itself regular WI.

\textbf{Commit-then-SWI:} Consider the following protocol
composing
  a statistically-binding commitment and SWI:

  \begin{description}
  \item [Common input:] $x\in L$ for an $\mathcal{NP}$-language $L$
  with corresponding $\mathcal{NP}$-relation $R_L$.
  \item [Prover auxiliary input:] $w$ such that $(x, w)\in R_L$.
  \item[The protocol:] consisting of two stages:
  \begin{description}
  \item [Stage-1:] The prover $P$ computes and sends $c_w=C(w,
  r_w)$, where $C$ is a statistically-binding commitment and $r_w$
  is the randomness used for commitment.

  \item [Stage-2:] Define a new language $L^{\prime}=\{(x, c_w)| \exists (w, r_w) \ s.t.\ c_w=C(w, r_w)
  \wedge R_L(x, w)=1  \}$. Then, $P$ proves to $V$ that it knows a
  witness to $(x, c_w)\in L^{\prime}$, by running a SWI protocol.

  \end{description}
  \end{description}

  One interesting observation for the above commit-then-SWI protocol
  is that commit-then-SWI is itself a regular WI for $L$.

  \begin{proposition} \label{CSWI}  Commit-then-SWI is itself a regular WI for
  the language $L$.
  \end{proposition}

  \textbf{Proof} (of Proposition \ref{CSWI}). For any PPT malicious
  verifier   $V^*$, possessing some auxiliary input $z\in \{0, 1\}^*$,
  and for any $x\in L$ and two (possibly different) witnesses $(w_0,
  w_1)$ such that $(x, w_b)\in R_L$ for both $b\in \{0, 1\}$,
  consider the executions of commit-then-SWI: $\langle P(w_0), V^*(z)\rangle (x)$ and
$\langle P(w_1), V^*(z)\rangle (x)$.

Note that for $\langle P(w_b), V^*(z)\rangle (x)$, $b\in \{0,
1\}$, the input to SWI of Stage-2 is $(x, c_{w_b}=C(w_b,
r_{w_b}))$, and the auxiliary input to $V^*$  at the beginning of
Stage-2 is $(x, c_{w_b}, z)$. Note that $(x, c_{w_0}, z)$ is
indistinguishable from $(x, c_{w_1}, z)$. Then, the regular WI
property of the whole composed  protocol is followed from the SWI
property of Stage-2. \hfill $\square$

\begin{definition}[system for argument/proof of knowledge \cite{G01,BG06}]
\label{POK} Let $R$ be a binary relation and $\kappa: N\rightarrow
[0, 1]$. We say that  a probabilistic polynomial-time (PPT)
interactive machine $V$ is a knowledge verifier for the relation
$R$ with knowledge error $\kappa$ if the following two conditions
hold:
\begin{itemize}
\item Non-triviality: There exists an interactive machine $P$ such
that for every $(x, w)\in R$ all possible interactions of $V$ with
$P$ on common input $x$ and auxiliary input $w$ are accepting.
\item Validity (with error $\kappa$): There exists a polynomial
$q(\cdot)$ and a probabilistic oracle machine $K$ such that for
every interactive machine $P^*$, every $x\in L_R$, and every $w,
r\in\{0, 1\}^*$, machine $K$ satisfies the following condition:

Denote by $p(x, w, r)$ the probability that the interactive
machine $V$ accepts, on input $x$, when interacting with the
prover specified by $P^*_{x, w, r}$ (where $P^*_{x, w, r}$ denotes
the strategy of $P^*$ on common input $x$, auxiliary input $w$ and
random-tape $r$). If $p(x, w, r)>\kappa(|x|)$, then, on input $x$
and with oracle access to $P^*_{x, w, r}$, machine $K$ outputs a
solution $w^{\prime}\in R(x)$ within an expected number of steps
bounded by
$$\frac{q(|x|)}{p(x, w, r)-\kappa (|x|)}$$ The oracle machine $K$ is
called a knowledge extractor.

\end{itemize}
An interactive argument/proof system $\langle P, V\rangle$  such
that $V$ is a knowledge verifier for a relation $R$ and $P$ is a
machine satisfying the non-triviality condition (with respect to
$V$ and $R$) is called a system for argument/proof of knowledge
(AOK/POK) for the relation $R$.
\end{definition}

The above definition of POK is with respect to
\emph{deterministic} prover strategy. POK also can be defined with
respect to \emph{probabilistic} prover strategy. It is recently
shown that the two definitions are equivalent for all natural
cases (e.g., POK for $\mathcal{NP}$-relations) \cite{BG06}.

We mention that  Blum's protocol for directed Hamiltonian Cycle
DHC \cite{B86} is just a 3-round public-coin WIPOK for
$\mathcal{NP}$, which is recalled below.

\textbf{Blum's protocol for DHC \cite{B86}.}  The $n$-parallel
repetitions of Blum's basic protocol
 for proving the knowledge of Hamiltonian cycle on a given directed graph $G$ \cite{B86} is just a 3-round
 public-coin WIPOK for
 $\mathcal{NP}$ (with knowledge error $2^{-n}$) under any one-way permutation (as the first round of it involves one-round
 perfectly-binding commitments of a random permutation of  $G$). But it can be easily modified into
 a 4-round public-coin WIPOK for $\mathcal{NP}$ under any OWF by
 employing  Naor's two-round (public-coin) perfectly-binding commitment
 scheme \cite{N91}. The following is the description of  Blum's \emph{basic} protocol for DHC:

\begin{description}
\item [Common input.] A directed graph $G=(V, E)$ with $q=|V|$
nodes.

\item [Prover's private input.] A directed Hamiltonian cycle $C_G$
in $G$.

\item [Round-1.] The prover selects a random permutation, $\pi$,
of the vertices $V$, and commits (using a perfectly-binding
commitment scheme)  the entries of the adjacency matrix of the
resulting permutated graph. That is, it sends a $q$-by-$q$ matrix
of commitments so that the $(\pi(i), \pi(j))^{th}$ entry is a
commitment to 1 if $(i, j)\in E$, and is a commitment to 0
otherwise.

\item [Round-2.] The verifier uniformly selects a bit $b\in \{0,
1\}$ and sends it to the prover.

\item [Round-3.] If $b=0$ then the prover sends $\pi$ to the
verifier along with the revealing of all commitments (and the
verifier checks that the revealed graph is indeed isomorphic to
$G$ via $\pi$); If $b=1$, the prover reveals to the verifier only
the commitments to entries  $(\pi(i), \pi(j))$ with $(i, j)\in
C_G$ (and the verifier checks that all revealed values are 1 and
the corresponding entries form a simple $q$-cycle).

\end{description}

 We remark that the WI property of Blum's protocol for DHC relies
 on the hiding property of the underlying perfectly-binding
 commitment scheme used in its first-round.

\textbf{Statistical WI argument/proof of knowledge (WIA/POK).} We
employ, in a critical way, constant-round \emph{statistical}
WIA/POK
 in this work. We briefly note two
simple ways for achieving statistical WIA/POK systems. Firstly,
for any statistical/perfect $\Sigma$-protocol (defined below), the
OR-proof (i.e., the $\Sigma_{OR}$-protocol) is statistical/perfect
WI proof of knowledge. The second approach is to modify the
(parallel repetition of) Blum's protocol for DHC \cite{B86} (that
is computational WIPOK) into constant-round statistical WIAOK by
replacing the statistically-binding commitments used in the
first-round of Blum's protocol by constant-round
\emph{statistically-hiding} commitments. One-round
statistically-hiding commitments can be based on any
collision-resistant hash function \cite{DPP93,HM96}. Two-round
statistically-hiding commitments can be based on  any claw-free
collection with an efficiently recognizable index set
\cite{GMR88,GK96,G01} (statistically-hiding commitments can also
be based on general assumptions, in particular any OWF, with
non-constant rounds \cite{NOVY98,HHK05,HR06}).

\subsection{$\Sigma$ and $\Sigma_{OR}$ Protocols}
$\Sigma$-protocols are very useful cryptographic tools that are
3-round public-coin protocols satisfying a special honest-verifier
zero-knowledge (SHVZK) property and a special soundness property
in the sense of knowledge extraction.

\begin{definition}[$\Sigma$-protocol \cite{C96}] A 3-round public-coin protocol $\langle P, V\rangle$ is said to
be a $\Sigma$-protocol for an $\mathcal{NP}$-language with
relation $R_L$ if the following hold:
\begin{itemize}
\item Completeness. If $P$, $V$ follow the protocol, the verifier
always accepts.

\item Special soundness. From any common input $x$ of length
$poly(n)$ and any pair of accepting conversations on input $x$,
$(a, e, z)$ and $(a, e^{\prime}, z^{\prime})$ where $e\neq
e^{\prime}$, one can efficiently compute $w$ such that $(x, w)\in
R_L$. Here $a$, $e$, $z$ stand for the first, the second and the
third message respectively and $e$ is assumed to be a string of
length $k$ (such that $1^k$ is polynomially related to the
security parameter  $1^n$) selected uniformly at random in $\{0,
1\}^k$.

\item   Special honest verifier zero-knowledge (SHVZK). There
exists a probabilistic polynomial-time (PPT) simulator $S$, which
on input $x$ (where there exists a $w$ such that $(x, w)\in R_L$)
and a random challenge string $\hat{e}$, outputs an accepting
conversation of the form $(\hat{a}, \hat{e}, \hat{z})$, with the
probability distribution that is indistinguishable from that of
the real conversation $(a, e, z)$ between the honest $P(w)$ and
$V$ on input $x$.
\end{itemize}
\end{definition}

A $\Sigma$-protocol is called \emph{perfect/statistical}
$\Sigma$-protocol, if it is perfect/statistical SHVZK. A
$\Sigma$-protocol is called \emph{partial witness-independent}, if
the generation of its first-round message is independent of (i.e.,
without using)   the witness for the common input. A very large
number of $\Sigma$-protocols have been developed in the
literature. In particular,  (the $n$-parallel repetition of)
Blum's protocol for DHC \cite{B86} is  a (partial
witness-independent) computational $\Sigma$-protocol  for
$\mathcal{NP}$; That is, the $n$-parallel repetition of Blum's
protocol for DHC \cite{B86} is  also a three-round  (partial
witness-independent) WI for $\mathcal{NP}$. Most practical
$\Sigma$-protocols for number-theoretical languages (e.g., DLP and
RSA \cite{S91,GQ88}, etc) are  (partial witness-independent)
\emph{perfect}
$\Sigma$-protocols. 
 For a good survey of $\Sigma$-protocols and
their applications, the reader is referred to \cite{D03}.


\textbf{$\Sigma$-Protocol for DLP \cite{S91}.}
 The following is a
$\Sigma$-protocol $\langle P, V\rangle$ proposed by Schnorr
\cite{S91} for proving the knowledge of discrete logarithm, $w$,
for a common input of the  form $(p, q, g, h)$ such that $h=g^w \
mod\ p$, where on  a security parameter $n$,
 $p$ is a uniformly selected
 $n$-bit prime such that  $q=(p-1)/2$ is also a prime,  $g$  is an element in $Z_p^*$ of order
$q$. It is also actually the first efficient $\Sigma$-protocol
proposed in the literature.

\begin{itemize}
\item $P$ chooses $r$ at random in $Z_q$  and sends $a=g^r \ mod \
p$ to $V$.

\item $V$ chooses a challenge $e$ at random in $Z_{2^k}$ and sends
it to $P$. Here, $k$ is fixed such that $2^k< q$.

\item $P$ sends $z=r+ew \ mod\  q$ to $V$, who checks that
$g^z=ah^e \ mod\ p$, that $p$, $q$ are prime and that $g$, $h$
have order $q$, and accepts iff this is the case.
\end{itemize}

 \textbf{The OR-proof of $\Sigma$-protocols
\cite{CDS94}.} 
One basic construction with $\Sigma$-protocols is the OR of a real
protocol conversation  and a simulated one, called $\Sigma_{OR}$,
that allows a prover to show that given two inputs $x_0$, $x_1$
(for possibly different $\mathcal{NP}$-relations $R_0$ and $R_1$
respectively), it knows a $w$ such that either $(x_0, w)\in R_0$
or $(x_1, w)\in R_1$, but without revealing which is the case
(i.e., witness indistinguishable WI) \cite{CDS94}.
 Specifically, given two
 $\Sigma$-protocols $\langle P_b, V_b\rangle$ for $R_b$, $b\in \{0, 1\}$,
with random challenges of, without loss of generality, the same
length $k$, consider the following protocol $\langle P, V\rangle$,
which we call $\Sigma_{OR}$. The common input of $\langle P,
V\rangle$ is $(x_0, x_1)$ and  $P$ has a private input $w$ such
that $(x_b, w)\in R_b$.

\begin{itemize}
\item $P$ computes the first message  $a_b$ in $\langle P_b,
V_b\rangle$, using $x_b$, $w$ as private inputs. $P$ chooses
$e_{1-b}$ at random,  runs the SHVZK simulator of $\langle
P_{1-b}, V_{1-b}\rangle$ on input $(x_{1-b}, e_{1-b})$, and lets
$(a_{1-b}, e_{1-b}, z_{1-b})$ be the output. $P$ finally sends
$a_0$, $a_1$ to $V$. \item  $V$ chooses a random $k$-bit string
$s$ and sends it to $P$. \item $P$ sets $e_b=s\oplus e_{1-b}$ and
computes the answer $z_b$ to challenge $e_b$ using $(x_b, a_b,
e_b,w)$ as input. He sends $(e_0, z_0, e_1, z_1)$ to $V$. \item
$V$ checks that $s=e_0 \oplus e_1$ and that conversations $(a_0,
e_0, z_o)$, $(a_1, e_1, z_1)$ are accepting conversations with
respect to inputs $x_0$, $x_1$, respectively.
\end{itemize}

\begin{theorem} \textup{\cite{CDS94}} The protocol $\Sigma_{OR}$ above is a $\Sigma$-protocol for $R_{OR}$, where
$R_{OR}=\{((x_0, x_1), w)|(x_0, w)\in R_0 \  \, \textup{or} \ \,
(x_1, w)\in R_1 \}$. Moreover, $\Sigma_{OR}$-protocols are witness
indistinguishable (WI) argument/proof of knowledge systems.
\end{theorem}

\textbf{The SHVZK simulator of $\Sigma_{OR}$ \cite{CDS94}.} For a
$\Sigma_{OR}$-protocol of the above form, denote by $S_{OR}$ the
perfect SHVZK simulator of it and denote by $S_b$ the perfect
SHVZK simulator of the protocol $\langle P_b, V_b\rangle$ for
$b\in\{0, 1\}$. Then on common input $(x_0, x_1)$ and a random
string $\hat{e}$ of length $k$, $S_{OR}((x_0, x_1), \hat{e})$
works as follows: It firstly chooses a random $k$-bit string
$\hat{e}_0$, computes $\hat{e}_1=\hat{e}\oplus \hat{e}_0$, then
$S_{OR}$ runs $S_b(x_b, \hat{e}_b)$ to get a simulated transcript
$(\hat{a}_b, \hat{e}_b, \hat{z}_b)$ for $b\in \{0, 1\}$, finally
$S_{OR}$ outputs $((\hat{a}_0, \hat{a}_1), \hat{e}, (\hat{e}_0,
\hat{z}_0, \hat{e}_1, \hat{z}_1))$.


\section{The BPK Model with Adaptive Language Selection}
 \label{APPBPK}

We present the  definitions of concurrent soundness and
concurrent zero-knowledge in the BPK model (cf.
\cite{CGGM00,MR01a,DV05,OPV06}). The key augmentation with the
current formulation, in comparison with previous definition of the
BPK model, is to allow adaptive language selection based on
public-keys. 

\subsection{Honest players in the BPK model}  We say a class of
languages $\mathcal{L}$ is \emph{admissible} to a protocol
$\langle P, V\rangle$ if the protocol can work (or, be
instantiated) for any language $L\in \mathcal{L}$. Typically,
$\mathcal{L}$ could be the set of all $\mathcal{NP}$-languages
(via $\mathcal{NP}$-reduction in case $\langle P, V\rangle$ can
work for an $\mathcal{NP}$-complete language) or the set of any
languages admitting $\Sigma$-protocols (in this case $\langle P,
V\rangle$ could be instantiated for any language in $\mathcal{L}$
efficiently without going through general
$\mathcal{NP}$-reductions).  Let $R_{KEY}$ be an
$\mathcal{NP}$-relation validating the public-key and secret-key
pair $(PK, SK)$ generated by honest verifiers, i.e., $R_{KEY}(PK,
SK)=1$ indicates that $SK$ is a valid secret-key of $PK$. Then, a
protocol $\langle P, V\rangle$ in the BPK model, w.r.t. some
admissible language set $\mathcal{L}$ and some key-validating
relation $R_{KEY}$, consists of the following:
\begin{itemize}
\item $F$, a public-key file that is a polynomial-size collection
of records ($id, PK_{id}$), where $id$ is a string identifying a
verifier and $PK_{id}$ is its (alleged) public-key. When
verifier's IDs are implicitly specified from the context, for
presentation simplicity we also just take $F$ as a collection of
public-keys in protocol specification and security analysis.

\item $\mathcal{M}$, a PPT language-selecting machine  that  on
inputs $(1^n, F)$ outputs  the description of an
$\mathcal{NP}$-relation $R_L$ for an $\mathcal{NP}$-language $L\in
\mathcal{L}$. The output of $\mathcal{M}$ (i.e., the description
of $R_L$) is then given to both the prover $P$ and  (proof-stage
of) the verifier
$V$. 
We require that given the description of $R_L$, the admissibility
of $L$ (i.e., the membership of  $L\in \mathcal{L}$) can be
efficiently decided.

\item $P(1^n, R_L,  x, w, F, id, \gamma)$, an honest prover that
is a polynomial-time interactive machine, where  $1^n$ is a
security parameter, $x$ is a $poly(n)$-bit string in $L$, $w$ is
an auxiliary input, $F$ is a public-file, $id$ is a verifier
identity, and $\gamma$ is its random-tape.

\item $V$,  an honest verifier that is a polynomial-time
interactive machine working in two stages.
\begin{enumerate}
\item Key generation stage. $V$, on a security parameter $1^n$ and
a random-tape $r$, outputs a  key pair $(PK, SK)$ satisfying
$R_{KEY}(PK, SK)=1$. $V$ then registers $PK$ in $F$ as its
public-key while keeping   the corresponding secret key $SK$ in
secret.

\item Proof stage. $V$, on inputs $SK$ and $R_L$, $x\in \{0,
1\}^{poly(n)}$ (which is supposed to be in $L$) and a random tape
$\rho$, performs an interactive protocol with a prover and outputs
``accept" indicating $x\in L$ or ``reject" indicating $x\not \in
L$.
\end{enumerate}
\end{itemize}

\textbf{Note:} 
On the one hand, augmenting the BPK model with adaptive language
selection complicates the formulation and may be more difficult to
fulfill  against adversaries with adaptive language selection
ability; but on the other hand, this is  a far more realistic
model for cryptographic protocols running concurrently in the
public-key model, where mixing the public-key structure as part of
the language is a natural adversarial strategy.

\subsection{The malicious concurrent prover and concurrent
soundness in the BPK model}  An $s$-concurrent malicious prover
$P^*$ in the BPK model, for a positive polynomial $s$, is a
probabilistic polynomial-time Turing machine that, on a security
parameter $1^n$ and  an auxiliary string $z\in \{0, 1\}^*$,
performs an $s$-concurrent attack against $V$ as follows in two
stages:

Let $(PK, SK)$ be the output of the key generation stage of $V$ on
a security parameter $1^n$ and a random string $r$. Then,  in the
first stage,   on inputs $(1^n, PK, z)$  $P^*$ first generates
$(R_L, \tau)$, where $R_L$ determines an \emph{admissible}
$\mathcal{NP}$-language $L\in \mathcal{L}$ and $\tau \in \{0,
1\}^*$
is some auxiliary information to be used in the second stage. %
We assume $P^*$ always selects an admissible language $L$ in the
first stage, otherwise the honest verifier will not start its
proof stages as we assume the admissibility of $L$ can be
efficiently verified. Then, in the second stage (i.e., proof
stage) w.r.t.
$R_L$ and $PK$, 
 $P^*$ can perform concurrently at most $s(n)$ interactive protocols (sessions)
with (the proof stage of) $V$ as follows:  If $P^*$ is already
running $i-1$ $(1\leq i \leq s(n))$ sessions, it can select
\emph{on the fly} a common input $x_i\in \{0, 1\}^{poly(n)}$
(which may be equal to $x_j$ for $1\leq j <i$) and initiate a new
session with the proof stage of $V(1^n, R_L, x_i, SK, \rho_i)$;
$P^*$ can output a message for any running protocol, and always
receive promptly  the response from $V$ (that is, $P^*$ controls
at its wish the schedule of the messages being exchanged in all
the concurrent sessions).
 We stress that in different sessions $V$ uses independent random-tapes in
its proof stage (that is,  $\rho_1, \cdots, \rho_{s(n)}$
 are independent random strings). We denote
by $view_{P^*}(1^n, z)$ the random variable describing the view of
$P^*$ in this experiment, which includes its random tape, the
auxiliary string $z$,  all messages it receives including the
public-key $PK$ and all messages sent by $V(1^n, R_L,  x_i, SK,
\rho_i)$'s in the $s(n)$ proof-stages, $1\leq i\leq s(n)$. For any
$(PK, SK)\in R_{KEY}$, we denote by $view^{V(SK)}_{P^*}(1^n, z,
PK)$ the random variable describing the view of $P^*$ specific to
$PK$, which includes its random tape, the auxiliary string $z$,
the (specific) $PK$, and  all messages it receives from  $V(1^n,
R_L, x_i, SK, \rho_i)$'s in the $s(n)$ proof-stages, $1\leq i\leq
s(n)$.

We then say a protocol $\langle P, V\rangle$ is \emph{concurrently
sound}  in the BPK model w.r.t. some admissible language set
$\mathcal{L}$, if for any sufficiently large $n$, for any honest
verifier $V$ and  all (except for a negligible fraction of) $(PK,
SK)$ outputted by  the key-generation stage of $V$,   for all
positive polynomials $s$ and all $s$-concurrent malicious prover
$P^*$ and any string $z\in \{0, 1\}^*$, for  any admissible
language $L\in \mathcal{L}$ and any string  $x \not \in L$ (of
length of $poly(n)$), the probability that $V$ outputs ``accept
$x\in L$"
in the $s$-concurrent attack against $V(1^n, R_L, SK)$ (i.e., in
one of the $s(n)$ sessions) is negligible in $n$, where the
probability is taken over the randomness of $P^*$, the randomness
of $V$ for key-generations and for all the $s(n)$ proof-stages.

\textbf{Notes:} The above concurrent soundness is defined w.r.t
multiple proof-stages (sessions) with the same public-key.
 In this case, we can imagine that the
auxiliary information $z$ encodes   information collected from
protocol executions w.r.t. other public-keys that are generated
independently of the public-key $PK$ at hand. 
Note that, as discussed in \cite{MR01a},  extension to the general
 case, where $P^*$ interacts with instances of multiple verifiers
 with multiple (independently generated) public-keys, is direct.
 Also note that all proof-stages of $V$ (i.e., all the $s(n)$
 sessions) are w.r.t. the same admissible language $L$. Such treatment is only for presentation simplicity. 
Both the security
 model and security proof of this work can be easily extended to the general
 case, where $P^*$ can select admissible language $L_i$ for
 each  session $i$, $1\leq i \leq s(n)$ (in this case, whenever $P^*$ starts a new session it sends $(x_i, R_{L_i})$
 to $V$ indicating that the new session is on common input $x_i$ and for admissible language $L_i$).

\subsection{The malicious concurrent verifier and concurrent ZK in
the BPK model} An  $s$-concurrent malicious verifier $V^*$, where
$s$ is a positive polynomial, is a PPT Turing machine that,  on
input $1^n$ and an auxiliary string $z$, works in two stages:
\begin{description}

\item [Stage-1 (key-generation stage).] On $(1^n, z)$ $V^*$
 outputs a relation $R_L$ determining an admissible  language $L\in \mathcal{L}$, an arbitrary public-file $F$ and a list of (without loss
of generality) $s(n)$ identities $id_1, \cdots, id_{s(n)}$. Then,
$V^*$ is given a list of $s(n)$   strings $\bar{\textbf{x}}=\{x_1,
\cdots, x_{s(n)}\} \in L^{s(n)}$ of  length $poly(n)$ each, where
$x_i$ might be equal to $x_j$, $1\leq i, j \leq s(n)$.


\item [Stage-2 (proof stage).] Starting from  the final
configuration of Stage-1, $V^*$ concurrently interacts with
$s(n)^2$ instances of the honest prover $P$: $P(1^n, F, R_L, x_i,
w_i, id_j, \gamma_{(i, j)})$, where $1\leq i, j \leq s(n)$, $(x_i,
w_i)\in R_L$  and $\gamma_{(i, j)}$'s are independent random
strings. In this stage, $V^*$ controls at its wish the schedule of
the messages being exchanged in all the concurrent sessions. In
particular, $V^*$ can output a message for any running session
dynamically based on the transcript up to now, and always receive
promptly  the response from
$P$. 
For any auxiliary string $z\in \{0, 1\}^*$, each public-key file
$F$ and $R_L$ outputted by $V^*$ in Stage-1 and any
$\bar{\textbf{x}}=\{x_1, \cdots, x_{s(n)}\} \in L^{s(n)}$, we
denote by $view_{V^*(z)}^{\{P(F, R_L, x_i, w_i, id_j, \gamma_{(i,
j)})'s \}}(1^n, \bar{\textbf{x}})$ the random variable describing
the view of $V^*$ in its second stage of this experiment, which
includes $(z, F, R_L, \bar{\textbf{x}})$, the randomness of $V^*$
in its second stage and all messages received from all the
$s(n)^2$ prover instances.
\end{description}

\begin{definition}  [concurrent zero-knowledge in the BPK model]  \label{CZK} A
protocol $\langle P, V\rangle$  is (black-box) concurrent
zero-knowledge in the BPK model w.r.t. some admissible language
set $\mathcal{L}$, if there exists a PPT black-box simulator $S$
such that for any sufficiently large $n$ and every $s$-concurrent
malicious verifier $V^*$ the following two
 distribution ensembles are indistinguishable:
  $$\{view_{V^*(z)}^{\{P(1^n, F, R_L, x_i, w_i, id_j,
\gamma_{(i, j)})'s \}}(1^n,
\bar{\textbf{x}})\}_{\bar{\textbf{x}}\in L^{s(n)}, L\in
\mathcal{L}, F\in \{0, 1\}^*,  z\in \{0, 1\}^*}$$
$$\{S(1^n, F, R_L, \bar{\textbf{x}}, z)\}_{\bar{\textbf{x}}\in
L^{s(n)}, L\in \mathcal{L}, F\in \{0, 1\}^*,  z\in \{0, 1\}^*}$$

\end{definition}

\textbf{Notes:} For presentation simplicity, the CZK property in
the BPK model with adaptive language selection is formulated with
respect to  that all $s(n)^2$ sessions (i.e., proof-stages) are
for the same $\mathcal{NP}$-relation $R_L$ and that
$\bar{\textbf{x}}\in L^{s(n)}$ are predefined (i.e., not selected
adaptively by $V^*$). Both the security model and security proof
of this work can be easily extended to the general cases, where
$V^*$ can select admissible language for each of the $s(n)^2$
sessions and can select the common inputs $x_i$'s adaptively. We
remark that for adaptive input selection, it is the responsibility
of $V^*$ to provide the corresponding $\mathcal{NP}$-witnesses
$w_i$'s to the honest prover instances.


\section{Motivation for Concurrent Knowledge-Extraction in the Public-Key Model}\label{appattack}

  We show  a concurrent interleaving and
malleating attack on  the concurrent ZK protocol of \cite{DV05}
that is both \emph{concurrently sound} and \emph{normal argument
of knowledge} in the BPK model, in which by concurrently
interacting with the honest verifier in two sessions a malicious
$P^*$ can (with probability 1)  malleate the verifier's
interactions in one session into successful interactions in
another session on a true (public-key related) statement but
without knowing any witness to the statement being proved. This
shows that concurrent soundness and normal arguments of knowledge
do not guarantee concurrent verifier security in the public-key
model. Actually, we show that, assuming any OWF,  CKE is
\emph{strictly} stronger than concurrent soundness in the
public-key model.  This
 serves a good motivation for understanding
``possession of knowledge on the Internet with registered
public-keys", i.e., the subtleties of concurrent
knowledge-extraction in the public-key
model.




\subsection{The Protocol Structure of \cite{DV05}}

\begin{description}
\item [Key-generation.] Let $f_V$ be a OWF that admits
$\Sigma$-protocols. On a security parameter $n$, each verifier $V$
 randomly selects two  elements in
the domain of $f_V$, $x^0_V$ and $x^1_V$ of length $n$ each,
computes $y^0_V=f_V(x^0_V)$ and $y^1_V=f_V(x^1_V)$.   $V$
publishes $(y^0_V, y^1_V)$
 as its public-key while keeping $x^b_V$ as its secret-key for a randomly chosen $b$  from $\{0,
 1\}$. (For OWF-based implementation, $V$ also publishes a random
 string $r_V$ of length $3n$ that serves the first-round message of
 Naor's OWF-based perfectly-binding commitment scheme \cite{N91}.)

\item [Common input.] An element $x\in L$ of length $poly(n)$,
where $L$ is an $\mathcal{NP}$-language that admits
$\Sigma$-protocols.

\item[The main-body of the protocol.] The main-body of the
protocol consists of the following three phases:
\begin{description}

\item[Phase-1.] The verifier $V$ proves to $P$ that it knows the
preimage of either $y^0_V$ or   $y^1_V$, by executing the
$\Sigma_{OR}$-protocol on  $(y^0_V, y^1_V)$
 in which $V$ plays the role of the knowledge prover. It is  additionally required that the first-round message of the $\Sigma_{OR}$-protocol
 is generated without using the preimage of either  $y^0_V$ or $y^1_V$ (i.e.,
 \emph{partial witness-independent}).
   Denote by $a_{V}$, $e_{V}$, $z_{V}$, the first-round, the second-round
and the third-round message of the $\Sigma_{OR}$-protocol of this
phase respectively. Here $e_{V}$ is the random challenge sent by
the prover to the verifier. (For OWF-based implementation, $P$
sends a random string $r_P$ of length $3n$ on the top, which
serves the first-round message of Naor's OWF-based
perfectly-binding commitments and  is used by $V$ in generating
$a_{V}$.)

 If $V$ successfully finishes the
 $\Sigma_{OR}$-protocol of this phase and $P$ accepts, then goto Phase-2.
 Otherwise, $P$ aborts.

 \item[Phase-2.] Let $TC$ be a trapdoor bit commitment scheme with the preimage of either $y^0_V$ or $y^1_V$
 as  the trapdoor. The prover randomly selects
 a string $\hat{e}\in \{0, 1\}^n$, and sends $c_{\hat{e}}=\{TCCom(\hat{e}_1), TCCom(\hat{e}_2), \cdots,
 TCCom(\hat{e}_n)\}$ to
 the verifier $V$, where $\hat{e}_i$ is the $i$-th bit of $\hat{e}$.

\item [Phase-3.] Phase-3 runs essentially the underlying
$\Sigma$-protocol for $L$ but with the random challenge set by a
coin-tossing mechanism. Specifically, the prover computes and
sends the first-round message of the underlying $\Sigma$-protocol,
denoted $a_P$, to the verifier $V$ (for OWF-based implementation,
$a_P$ is computed also using $r_V$ published by $V$ in the
key-generation phase); Then $V$ responds with a random challenge
$q$; Finally, $P$ reveals $\hat{e}$ (committed in Phase-2), sets
$e_P=\hat{e}\oplus q$, and computes the third-round message of the
underlying $\Sigma$-protocol for $L$, denoted $z_P$, with $e_P$ as
the real random challenge.

\item[Verifier's decision.] $V$ accepts if and only if  $\hat{e}$
is decommitted correctly and $e_P=\hat{e}\oplus q$ and $(a_P, e_P,
z_P)$ is an accepting conversation for $x\in L$.

\end{description}
\end{description}

\textbf{Remark:} The above protocol structure is essentially that
of the incomplete CZK protocol of \cite{Z03} (Figure-3, page 17),
and can be implemented based on any OWF. The key difference  in
the actual implementations of \cite{Z03,DV05} is that \cite{DV05}
uses a special trapdoor commitment scheme in Phase-2, where the
decommitment formation to 0 or 1 is in turn committed in  two
statistically-binding commitments. This technique is critical for
achieving concurrent soundness, the reader is referred to
\cite{DV05} for more details. We remark that the differences in
actual implementations do not invalidate  the  attack presented
below  in Section \ref{attack}, which is presented with respect to
a more general protocol structure.


\subsection{The concurrent interleaving and malleating attack }
\label{attack}

With respect to the above protocol structure of  the protocols of
\cite{DV05,Z03}, 
 let $\hat{L}$ be any $\mathcal{NP}$-language admitting a
$\Sigma$-protocol that is denoted by $\Sigma_{\hat{L}}$ (\emph{in
particular, $\hat{L}$ can be an empty set}). Then for an honest
verifier $V$ with its public-key $PK=(y^0_V, y^1_V)$, we define a
new language $L=\{(\hat{x}, y^0_V, y^1_V)| \exists w \ s.t. \
(\hat{x}, w)\in R_{\hat{L}} \ \text{OR}\ y^b_V=f_V(w) \ for \ b
\in \{0, 1\}\}$. Note that for any string $\hat{x}$ (whether
$\hat{x}\in \hat{L}$ or not), the statement ``$(\hat{x}, y^0_V,
y^1_V) \in L$'' is always true as $PK=(y^0_V, y^1_V)$ is honestly
generated. Also note that $L$ is a language that admits
$\Sigma$-protocols (as $\Sigma_{OR}$-protocol is itself a
$\Sigma$-protocol). Now, we describe the concurrent interleaving
and malleating attack,  in which $P^*$ successfully convinces the
honest verifier of the statement ``$(\hat{x}, y^0_V, y^1_V) \in
L$'' for \emph{any arbitrary} $poly(n)$-bit string $\hat{x}$
(\emph{even when $\hat{x}\not \in \hat{L}$}) by concurrently
interacting with $V$ in two sessions as follows.

\begin{enumerate}
\item $P^*$ initiates the first session with $V$. (For OWF-based
implementation, $P$ just sends $r_P=r_V$ as its first message to
$V$, where $r_V$ is the random string registered by $V$ as a part
of its public-key for OWF-based implementation.) After receiving
the first-round message, denoted by  $a^{\prime}_V$, of the
$\Sigma_{OR}$-protocol  of Phase-1 of the first session on common
input $(y^0_V, y^1_V)$ (i.e.,  $V$'s public-key), $P^*$ suspends
the first session.

\item  $P^*$ initiates a second session   with $V$,  and works
just as the honest prover does in Phase-1 and Phase-2 of the
second session. We denote by $c_{\hat{e}}$ the Phase-2 message of
the second session (i.e., $c_{\hat{e}}$ commits to a random string
$\hat{e}$ of length $n$). When $P^*$ moves into Phase-3 of the
second session and needs to send $V$ the first-round message,
denoted by $a_P$, of the $\Sigma$-protocol of Phase-3 of the
second session \emph{on common input $(\hat{x}, y^0_V, y^1_V)$},
$P^*$ does the following:

\begin{itemize}

\item $P^*$ first runs the  SHVZK simulator of $\Sigma_{\hat{L}}$
(i.e.,  the $\Sigma$-protocol for $\hat{L}$) on $\hat{x}$ to get a
simulated conversation, denoted by $(a_{\hat{x}}, e_{\hat{x}},
z_{\hat{x}})$, for the (\emph{possibly false})  statement
``$\hat{x} \in \hat{L}$".

\item $P^*$ sets $a_P=(a_{\hat{x}}, a^{\prime}_V)$ and sends $a_P$
to $V$ as the first-round message of the $\Sigma$-protocol of
Phase-3 of the second session, where $a^{\prime}_V$ is the one
received by $P^*$ in the first session.

\item After receiving the second-round message of Phase-3 of the
second session, denoted by $q$ (i.e., the random challenge from
$V$), $P^*$ sets $e_P=\hat{e}\oplus q$ and then suspends the
second session.
\end{itemize}

\item $P^*$ continues the first session, and sends
$e^{\prime}_V=\hat{e}\oplus q\oplus e_{\hat{x}}=e_P\oplus
e_{\hat{x}}$ as the second-round message of the
$\Sigma_{OR}$-protocol of Phase-1 of the first session.

\item After receiving the third-round message of the
$\Sigma_{OR}$-protocol of Phase-1 of the first session, denoted by
$z^{\prime}_V$, $P^*$ suspends the first session again.

\item $P^*$ continues the execution of the  second session again,
reveals $\hat{e}$ committed in Phase-2 of the second session, and
sends to $V$ $z_P=((e_{\hat{x}}, z_{\hat{x}}), (e^{\prime}_V,
z^{\prime}_V))$ and the decommitment information of $\hat{e}$  as
the last-round message of the second session.

\end{enumerate}

 Note that  $(a_{\hat{x}}, e_{\hat{x}},
z_{\hat{x}})$ is an accepting conversation for the (possibly
false) statement ``$\hat{x}\in \hat{L}$",  $(a^{\prime}_V,
e^{\prime}_V, z^{\prime}_V)$ is an accepting conversation for
showing the knowledge of the preimage of either $y^0_V$ or
$y^1_V$, and furthermore $e_{\hat{x}}\oplus
e^{\prime}_V=e_P=\hat{e}\oplus q$. According to the description of
$\Sigma_{OR}$ (presented in Section  \ref{basic}), this means
that, from the viewpoint of $V$, $(a_P, e_P, z_P)$ is an accepting
conversation  of Phase-3 of the second-session on common input
$(\hat{x}, y^0_V, y^1_V)$. That is, $P^*$ successfully convinced
$V$ of the statement ``$(\hat{x}, (y^0_V, y^1_V)) \in L$''
(\emph{even for $\hat{x}\not \in \hat{L}$}) in the second session
  \emph{but without knowing any
corresponding $\mathcal{NP}$-witness}! This demonstrates that the
 protocol of \cite{DV05} fails to be a proof of knowledge (fails knowledge
extraction) in concurrent executions (note that it was not
designed as such, since this new issue is the notion we put forth
here). We remark that mixing the public key structure as part of
the language is a natural attack strategy for the public-key model
(a different demonstration of this was given in \cite{YZ06}).


\section{Formulating Concurrent Knowledge-Extraction in the
Public-Key Model} \label{CKEformulation}

Now, we proceed to formulate concurrent verifier security in light
of the above concrete attack against the protocol of
\cite{Z03,DV05}. Note that the concrete attack is of
\emph{man-in-the-middle} (MIM) nature, and is related to
malleability of protocols.   The security notion  assuring that  a
malicious  prover $P^*$ does ``know" what it claims to know, when
it is concurrently interacting with the honest verifier $V$,  can
informally be formulated as:
%
%
%
 for any $x$,  if $P^*$ can convince $V$ (with public-key
$PK$) of ``$x\in L$'' (for an $\mathcal{NP}$-language $L$)  by
concurrent interactions, then there exists a 
 PPT  knowledge-extractor that outputs a witness
for $x\in L$. This is a natural extension of the normal arguments
of knowledge into the concurrent settings in the public-key model.
However, such a definition  does \emph{not} work in the public-key
model. The reason is:  the statements being proved may be related
to $PK$, and thus the extracted witness may be related to its
corresponding secret-key  $SK$ (actually, for the malicious prover
strategy of the concrete attack on the protocol of
\cite{Z03,DV05}, the extracted witness will just be the same
secret-key used by the knowledge-extractor); But, in
knowledge-extraction the PPT extractor may have already possessed
$SK$. To solve this subtlety, we require the extracted witness,
together with adversary's view,  to be \emph{independent} of $SK$.
But, the problem here is how to formalize such independence, in
particular, w.r.t. 
 a concurrent MIM? We solve this
in the spirit of non-malleability formulation \cite{DDN00}. That
is, we consider the message space (distribution) of $SK$, and such
independence is roughly  formulated as follows:
%
%
let $SK$ be the  secret-key and $SK^{\prime}$ is an element
randomly and independently distributed over the space of $SK$,
then we require that, for any polynomial-time computable relation
$R$, the probability $\Pr[R(\bar{w}, SK, view)=1]$ is negligibly
close to $\Pr[R(\bar{w}, SK^{\prime}, view)=1]$, where $\bar{w}$
is the set of  witnesses extracted by
the knowledge extractor for successful concurrent sessions and $view$ is the view of the adversary $P^*$. This captures the intuition 
 that $P^*$ does, in fact,  ``know'' the witnesses to the statements whose validations are successfully
 conveyed by concurrent interactions. 
%

\begin{definition}  [concurrent knowledge-extraction (CKE) in the public-key
model]\label{cke}
  We  say that  a protocol $\langle P, V\rangle$ is \textsf{\textup{concurrently
  knowledge-extractable}}  in the BPK  model w.r.t. some admissible language set $\mathcal{L}$ and some
  key-validating relation  $R_{KEY}$, if for  any positive polynomial $s(\cdot)$,
any $s$-concurrent malicious prover $P^*$ defined in Section
\ref{basic}, there exist a pair of (expected) polynomial-time
algorithms $S$ (the simulator) and $E$ (the extractor) such that
for any sufficiently large $n$, any auxiliary input $z \in \{0,
1\}^*$, and any polynomial-time computable relation $R$ (with
components drawn from $\{0, 1\}^* \cup \{\perp\}$), the following
hold, in accordance with the experiment
\textbf{$\textsf{Expt}_{\textup{CKE}}(1^n, z)$}  described below
(page \pageref{experiment}):


 \begin{table}[!h]
\begin{center}

\begin{tabular} {|c|}
 \hline \label{experiment}

 \textbf{$\textsf{Expt}_{\text{CKE}}(1^n, z)$}\\ 

\begin{minipage}[t] {6.0in} \small
\vspace{0.1cm}

\textbf{The simulator $S=(S_{KEY}, S_{PROOF})$:}

$(PK, SK, SK^{\prime})\longleftarrow S_{KEY}(1^n)$, where the
distribution of $(PK, SK)$ is identical with that of the output of
the key-generation stage of the honest verifier $V$,  $R_{KEY}(PK,
SK)=R_{KEY}(PK, SK^{\prime})=1$ and the distributions of $SK$ and
$SK$ are identical and \emph{independent}. 
In other words, $SK$ and $SK^{\prime}$ are two random and
independent secret-keys corresponding to $PK$. \\

 $(str, sta) \longleftarrow S_{PROOF}^{P^*(1^n,\ PK,\  z)}(1^n,  PK,  SK,
z)$. That is,   on inputs $(1^n, PK, SK, z)$ and with oracle access
to $P^*(1^n, PK, z)$, the simulator $S$ outputs a simulated
transcript $str$,   and some state information $sta$  to be
transformed to the knowledge-extractor $E$.\\

We denote by $S_1(1^n, z)$ the random variable  $str$ (in accordance
with above processes of $S_{KEY}$ and $S_{PROOF}$). For any $(PK,
SK)\in R_{KEY}$ and any $z\in \{0, 1\}^*$, we denote by $S_1(1^n,
PK, SK, z)$ the random variable describing the first output of
$S_{PROOF}^{P^*(1^n,\ PK,\ z)}(1^n, PK,  SK, z)$ (i.e., $str$
specific to $(PK, SK)$).
\\

\textbf{The knowledge-extractor $E$:}

$\overline{w} \longleftarrow E(1^n, sta, str)$. On $(sta, str)$, $E$
outputs a list of witnesses to statements whose validations are
successfully  conveyed in $str$.

\end{minipage}
\\
\hline
\end{tabular}
\end{center}

\end{table}

\begin{itemize}
\item \textbf{\textup{ Simulatability.}}
The following ensembles are identical (or indistinguishable): \\
 $\{S_1(1^n, PK, SK, z)\}_{(PK, SK)\in R_{KEY}, z\in\{0, 1\}^*}$ and
   $\{view^{V(SK)}_{P^*}(1^n, z, PK) \}_{(PK, SK)\in R_{KEY}, z\in\{0,
  1\}^*}$ (defined in Section
\ref{basic}). This in particular  implies that $str$ includes
$(PK, z)$, and
   the probability ensembles
$\{S_1(1^n, z)\}_{z\in\{0, 1\}^*}$ and $\{P^*(1^n, z)\}_{z\in\{0,
1\}^*}$ (defined in Section \ref{basic}) are actually identical
(or indistinguishable).


%



\item \textbf{\textup{Secret-key independent knowledge-extraction.}}  $E$, on inputs  $(1^n, str, sta)$,
 outputs witnesses to all statements successfully proved in
accepting  sessions in $str$. Specifically, $E$ outputs a list of
strings $\overline{w}=(w_1, w_2, \cdots, w_{s(n)})$, satisfying the
following:

\begin{itemize}
\item $w_i$ is set to be $\perp$, if the $i$-th  session in $str$ is not
accepting (due to abortion or verifier verification failure), where
$1\leq i \leq s(n)$.

\item \textsf{\textup{Correct knowledge-extraction for (individual) statements:}}
In any other cases (i.e., for  successful sessions), with
overwhelming probability $(x_i, w_i)\in R_L$,  where $x_i$ is the
common input selected by $P^*$ for the $i$-th session in $str$ and
$R_L$ is the admissible $\mathcal{NP}$-relation for $L\in
\mathcal{L}$ set by $P^*$ in $str$.

\item \textsf{\textup{ (Joint) knowledge extraction independence (KEI):}} \
$\Pr[R(SK, \overline{w}, str)=1]$ is negligibly close to
$\Pr[R(SK^{\prime}, \overline{w}, str)=1]$. 


\end{itemize}
The probabilities are taken over the randomness of $S$ in the
key-generation stage (i.e., the randomness for generating $(PK, SK,
SK^{\prime})$) and in all proof stages,  the randomness of $E$, and
the randomness of $P^*$. If the KEI property holds for any (not
necessarily polynomial-time computable) relation $R$, we say the
protocol $\langle P, V\rangle$ satisfies  \emph{statistical} CKE and
\emph{statistical} KEI.
\end{itemize}

\end{definition}


\subsection{Discussion and justification of the CKE
formulation}\label{CKEdiscussion}

We first note that the above CKE formulation follows the
simulation-extraction approach of \cite{PR05f} (which is also used
in \cite{BPS06}). Here, the key augmentation, besides some other
 adaptations  in the public-key model,  is the
property of \textsf{knowledge-extraction independence (KEI)}
explicitly required. Though the CKE and KEI notions are formulated
in the framework of public-key model, they are actually applicable
to protocols in the plain model,  in general, in order to capture
knowledge extractability against concurrent adversaries
interacting
with honest players of  secret values. 

\textbf{Simulated public-keys vs. real public-keys.} In our CKE
formulation, the simulation-extraction is w.r.t. \emph{simulated}
public-keys. In this case, explicitly requiring the KEI property
is crucial  for correctly formulating CKE, as the
simulator/extractor possesses the secret-keys corresponding to the
simulated public-keys. A natural and intuitive strengthening of
the CKE formulation might be: the simulator/extractor uses the
\emph{same}
public-keys of the  honest verifiers.  
  Specifically, for any concurrent malicious $P^*$ there
exists a PPT simulator/extractor that, \emph{on the same
public-key of the  honest verifier},  outputs  a simulated
transcript (that is indistinguishable from the real  view of
$P^*$) together with all witnesses to accepting sessions. In this
case, as the simulator/extractor does not possesses the secret-key
(of the honest verifier), the KEI property can be waived. But, the
key observation here is: constant-round CKE (\emph{whether ZK or
not}) with real public-keys are impossible. Specifically,
constant-round CKE with real public-keys implies constant-round
CZK (actually, potentially concurrently non-malleable ZK proof of
knowledge) \emph{in the plain model} by viewing verifier's
public-keys as a part of common inputs, which is however
impossible at least in the
black-box sense \cite{CKPR02}.  


\textbf{On the non-triviality  of KEI even with independent
languages.} With the above CKE formulation, we are actually
formulating  the independence of the witnesses, used
(``\emph{known}") by concurrent MIM adversary,  on the secret-key
(witness)  used by verifier (who may in turn play the role of
prover in some sub-protocols). A naive solution for KEI, which
appears to make sense in certain scenarios, may be to require the
language and statements being proved  are independent of
verifier's public-keys. But,  this way does not work in general.
Firstly note that, if the protocol is for $\mathcal{NP}$-Complete,
the statements being proved, selected adaptively by the adversary,
can be always related to verifier's public-key (e.g., via
$\mathcal{NP}$-reductions); Moreover, 
 for protocols in the BPK model, verifier's keys are used in essential ways,
particularly  in order to achieve round efficiency. This is the
case, especially when the protocol in the public-key model runs
concurrently over Internet (note that most concurrently secure
cryptographic tasks cannot be implemented
round-efficiently in the plain model). 
Typically,   a constant-round cryptographic protocol in the BPK
model  consists of several sub-protocols, such that the common
statement and verifier's public-keys  are mixed into  the inputs
to some sub-protocols. In this case,  even if the language (and
even if the witness being used by the \emph{honest} prover) is
independent of verifier's public-keys,   the inputs to the
sub-protocols, \emph{selected and decided by the concurrent
adversary based on its view of  concurrent interleaving attacks},
can be always related to (dependent on) verifier's keys (a typical
illustration is the Feige-Shamir-ZK-like protocols in the
public-key model \cite{F90}). The various concurrent interleaving
and malleating attacks presented in this work (in particular,  the
attack against the protocol variant of the efficient CZK-CKE
without $c_{sk}$ in Section \ref{attackwithoutcsk}) just
demonstrate such cases.

\textbf{CKE vs. concurrent soundness.} We show that, assuming any
OWF,  CKE is a strictly stronger notion for concurrent verifier
security than concurrent soundness in the public-key model.

\begin{proposition}\label{CSCKE}
Assuming any OWF, CKE is strictly stronger than concurrent
soundness in the public-key model.
\end{proposition}

 \textbf{Proof.} (of Proposition \ref{CSCKE}) \
 It's  easy to see that CKE  implies concurrent
soundness  in the public-key model. Specifically, suppose that for
some $(PK, SK)\in R_{KEY}$,  some admissible language $L$ and some
string $x\not\in L$ $P^*$ can convince $V(R_L, SK)$  of the false
statement ``$x\in L$"  with non-negligible probability in real
execution, then with almost the same probability (up to a
negligible gap) $P^*$ can convince the simulator $S(R_L, SK)$  of
$x\in L$  in $\textsf{Expt}_{\text{CKE}}(1^n, z)$ by the property
of simulatability, which however contradicts the secret-key
independent
knowledge-extraction property. 

Then the proposition is direct from the
 attack demonstrated in
Section \ref{attack} on the CZK protocol of \cite{DV05} that is
both concurrently sound and normal argument of knowledge and can
be implemented based on any OWF.
 Specifically, for the specific
strategy of $P^*$ of the  concurrent interleaving and malleating
attack, suppose $\hat{x}\not\in \hat{L}$ or just $\hat{L}$ is
empty, the witness extracted by any polynomial-time
knowledge-extraction algorithm $E$ (with $SK=x^b_V$ as its input)
must be the preimage of either $y^0_V$ or $y^1_V$. But, according
to the one-wayness of $f_V$ used in the key-generation stage, with
overwhelming probability the extracted witness will be the
preimage of $y^b_V$ conditioned on $E$ outputs a witness.
(Specifically, consider the simulator/extractor  emulates the
key-generation of the honest verifier, except that the value
$y^{1-b}_V$ is received externally as its input.) Define the
relation $R$ as: $R(w, SK, \cdot)=1$ if $f_V(w)=f_V(SK)$. Then,
conditioned on $E$ outputs a witness, the extracted witness (i.e.,
the preimage of $y^b_V$) is always related to $SK=x^b_V$, but can
be related to a random and independent $SK^{\prime}$ with
negligible probability. Thus, the CZK protocol of \cite{DV05} is
not concurrently knowledge-extractable in the public-key model.
\hfill $\square$







\section{Generic CZK-CKE 
 in the BPK Model} \label{genericCKE}

In this section, we present the generic constant-round  CZK-CKE
arguments for $\mathcal{NP}$ in the BPK model under standard
hardness  assumptions. The starting point is the basic and famous
Feige-Shamir ZK (FSZK) structure \cite{F90}. The FSZK structure is
conceptually simple, which simply composes two WIPOK
sub-protocols. In more details, let $f$ be a OWF, in the first
WIPOK sub-protocol with the verifier $V$ serving as the
knowledge-prover, $V$ computes $(y_0=f(s_0), y_1=f(s_1))$ for
randomly chosen $s_0$ and $s_1$; then $V$ proves to the prover $P$
the knowledge of the preimage of either $y_0$ or $y_1$. In the
second WIPOK sub-protocol with $P$ serving as the
knowledge-prover, on common input $x$, $P$ proves to $V$ the
knowledge of either a valid $\mathcal{NP}$-witness $w$ for $x\in
L$ or the preimage of either $y_0$ or $y_1$. FSZK is also argument
of knowledge, and can be high practically instantiated (without
going through general $\mathcal{NP}$-reductions) by the
$\Sigma_{OR}$ technique \cite{CDS94}. 

Let $(y_0, y_1)$ serve as the public-key of $V$ and $s_b$ (for a
random bit $b$) as the secret-key, the public-key version of FSZK
is CZK  in the BPK model. But, we shew  that the public-key
version of FSZK is not concurrently sound \cite{YZ06}, needless to
say concurrent knowledge-extractability (indeed, FSZK was not
designed for the  public-key model). We hope to add the CKE
property to FSZK in the BPK model (and thus get concurrent
security both for the prover and for the verifier simultaneously),
while remaining its conceptual simple structure as well as the
ability of practical instantiations.

The subtle point here is: we are actually  facing (dealing with) a
concurrent MIM 
 (CMIM), who manages  to malleate,
in a malicious and unpredictable way, the public-keys and
knowledge-proof interactions of the verifier in one session into
the statements and knowledge-proof interactions in  another
concurrent session. To add CKE security to FSZK in the BPK model,
some non-malleable (maybe inefficient) building tools seem to be
intrinsically required. In this work, we show how to do so without
 employing any non-malleable building tools.

The idea is to strengthen the first sub-protocol to be
\emph{statistical} WIPOK, and require the prover to first commit,
before starting the second WI sub-protocol, the supposed witness
to $c_w$ by running a \emph{statistically-binding} commitment
scheme $C$.
 This guarantees that if the witness committed to $c_w$ is dependent on the secret-key used by $V$,
there are indeed some differences between the interaction
distribution when $V$ uses $SK=s_0$ and that 
  when $V$ uses $SK=s_1$, and we can use such
distribution differences to violate the statistical WI of the
first sub-protocol. But, this solution loses CZK in general, as
the second WI sub-protocol is run w.r.t. commitments to different
values in real interactions and in the simulation. This problem
can be got passed by using a stronger second sub-protocol, i.e.,
the  strong WI (SWI) \cite{G01}.
 Note that the composition of
commitment  and SWI is itself regular WI, and thus CZK property is salvaged.  

The generic construction is depicted in Figure \ref{CKE-generic},
page \pageref{CKE-generic} (as the generic construction is for
$\mathcal{NP}$ via $\mathcal{NP}$-reduction,  we do not explicitly
describe  the language-selecting machine $\mathcal{M}$ in the
protocol specification).

\begin{center}
 \begin{figure}[!t]
\begin{tabular} {|c|}
 \hline
\begin{minipage}[t]{6.4in} \small \textbf{Key Generation.}  Let $f: \{0, 1\}^n\rightarrow \{0, 1\}^n$
be any OWF,  where $1^n$ is the system security parameter. Each
verifier $V$ selects
 random strings   $s_0$, $s_1$ from $\{0, 1\}^n$, randomly selects a bit $b\leftarrow\{0, 1\}$,
 computes $y_b=f(s_b)$ and sets $y_{1-b}=f(s_{1-b})$. $V$  registers $PK=(y_0, y_1)$ in a public file $F$ as its public-key, and keeps $SK=s_b$
 as its secret-key. Define $R_{KEY}=\{((y_0, y_1), s)|y_0=f(s) \vee y_1=f(s) \}$


    \end{minipage}\\ \hline

\begin{minipage}[t]{6.4in} \small
\begin{description}
\item [Common input.] An element $x\in L\cap\{0, 1\}^{poly(n)}$,
where $L$ is an $\mathcal{NP}$-Complete language with  the
corresponding $\mathcal{NP}$-relation  $R_L$.

\item [P private input.] An $\mathcal{NP}$-witness $w\in \{0,
1\}^{poly(n)}$ for $x\in L$. Here, we assume w.l.o.g. that the
witness for any $x\in L\cap\{0, 1\}^{poly(n)}$ is of the same
length $poly(n)$. 
\end{description}
\end{minipage}\\ \hline


\begin{minipage}[t]{6.4in} \small

\begin{description}
\item \item [Stage-1.] $V$ proves to $P$ that it knows a preimage
to  one of $y_0, y_1$, by running a \emph{statistical}  WIA/POK
protocol for $\mathcal{NP}$,  in which $V$ plays the role of
knowledge prover. The witness used by $V$ in this stage is $s_b$.

\item [Stage-2.] If $V$ successfully finishes Stage-1, $P$ does
the following: it computes and sends $c_w=C(w, r_w)$, where $C$ is
a statistically-binding commitment scheme and $r_w$ is the
randomness used for commitments.

\item [Stage-3.] Define a new $\mathcal{NP}$-language
$L^{\prime}=\{(x, y_0, y_1, c_w)| (\exists (w, r_w)\  s.t. \ c_w=
C(w, r_w) \wedge ( (x, w)\in R_L) \vee y_0=f(w) \vee y_1=f(w))\}$.
Then, $P$ proves to $V$ that it knows a witness for $(x, y_0, y_1,
c_w)\in L^{\prime}$, by running a strong WI argument/proof of
knowledge (WIA/POK) protocol for $\mathcal{NP}$. 

\end{description}
\end{minipage}


\\
\hline
\end{tabular}
\caption{\label{CKE-generic}   The generic CZK-CKE argument
$\langle P, V \rangle$ for $\mathcal{NP}$ in the BPK model}
\end{figure}
\end{center}

\subsection{Security analysis}\label{analysis}


\textbf{Notes on the underlying hardness assumptions and
round-complexity.} If the OWF $f$ used in key-generation admits
perfect/statistical $\Sigma$-protocols (and thus we can use
$\Sigma_{OR}$ in Stage-1), and we use Feige-Shamir ZK (FSZK) of
\cite{FS89} (with WI is replaced by $\Sigma_{OR}$)  to replace SWI
of Stage-3, the protocol depicted in Figure \ref{CKE-generic} can
be based on any OWF admitting perfect/statistical
$\Sigma$-protocols, and be of optimal (i.e., 4-round)
round-complexity by  round combinations; If we use in Stage-1 the
modified Blum's protocol for DHC with constant-round
statistically/perfectly hiding commitments, the protocol depicted
in Figure \ref{CKE-generic} can be based on any
collision-resistant hash function or any claw-free collection with
efficiently recognizable index set.


\begin{theorem} The protocol depicted in Figure \ref{CKE-generic} is  a constant-round  concurrently
knowledge-extractable concurrent ZK (CZK-CKE)  argument for
$\mathcal{NP}$ in the BPK  model.
\end{theorem}

\textbf{Proof.} 
 The completeness of the protocol
$\langle P, V\rangle$ can be easily checked.

\textbf{Concurrent zero-knowledge.}

We first consider a mental simulator $M$ that takes as input all
secret-keys corresponding to all public-keys registered in the
public-key file, \emph{in case the corresponding secret-keys
exist}.

For any $s(n)$-concurrent malicious verifier $V^*$ (defined in
Section \ref{APPBPK}) and any $\mathcal{NP}$-language $L$, $M$
runs $V^*$ as a subroutine on inputs $\bar{\textbf{x}}=\{x_1,
\cdots, x_{s(n)}\}\in L^{s(n)}$ (where $x_i$ might equal $x_j$,
$1\leq i, j \leq s(n)$ and $i\neq j$), the public file $F=\{PK_1,
\cdots, PK_{s(n)}\}$ and all assumed existing secret-keys. $M$
works just as the honest prover does in Stage-1 of any session. In
Stage-2 of any session on a common input $x_i$ and with respect to
a public-key $PK_j$ (i.e., the $i$-th session w.r.t $PK_j$, $1\leq
i, j \leq s(n)$),  $M$ computes $c^{(i)}_w=C(SK_j,  r^{(i)}_w)$,
where $SK_j$ is the secret-key corresponding to $PK_j$  for which
we assume it exists and $M$ knows. Then, on input $(x_i, PK_j,
c^{(i)}_w)$ $M$ runs the strong WI argumnet/proof of knowledge for
$\mathcal{NP}$ in Stage-3 of the session with $(SK_j,
r^{(i)}_{w})$ as its witness.

Then, by a simple hybrid argument, the indistinguishability
between the output of $M$ and the view of $V^*$ in real concurrent
interactions is direct from the regular WI of commit-then-SWI.
Note that, as mentioned in Section \ref{basic},  regular WI
preserves under concurrent composition in this case.

Finally, to build a PPT simulator $S$  from scratch, where $S$
does not know any secret-keys corresponding to public-keys in the
public file, we resort to the
 technique developed in \cite{CGGM00}.
Specifically, $S$ works in $s(n)+1$ phases. In each phase, $S$
either successfully finishes the simulation, or ``covers"  a  new
public-key  for which it has not known the corresponding
secret-key up to now  in case $V^*$ successfully finishes the
Stage-1 interactions w.r.t. that public-key.  Key coverage  is
guaranteed by the POK property of Stage-1 interactions. For more
details, see \cite{CGGM00,L01}.

\textbf{(\emph{Statistical}) concurrent knowledge-extraction.}

According to the CKE formulation, for any $s$-concurrent malicious
prover $P^*$ (defined in Section \ref{basic}) we need to build two
algorithms $(S, E)$. The simulator $S$, on inputs $(1^n, z)$,
works as follows: It first perfectly emulates the key-generation
stage of the honest verifier, getting $PK=(y_0, y_1)$ and $SK=s_b$
and $SK^{\prime}=s_{1-b}$ for a random bit $b$. Then, $S$ runs
$P^*$ on $(1^n, PK, z)$ to get $(R_L, \tau)$, where $R_L$
indicates an $\mathcal{NP}$-language for which the proof-stages
will work and $\tau$ is some auxiliary information to be used by
$P^*$ in proof-stages. In the proof stages, $S$ perfectly emulates
the honest verifier with the secret-key $SK$. Finally, whenever
$P^*$ stops, $S$ outputs the simulated transcript $str$, together
with the state information $sta$ set to be $(PK, SK, SK^{\prime},
z)$ and the random coins used by $S$.  Note that the simulated
transcript $str$ is identical to the view  of $P^*$ in real
execution.

The knowledge-extraction process is similar to that of
\cite{PR05f}. Note that we need to extract witnesses to all
accepting sessions in $str$.  Given $(str, sta)$, the
knowledge-extractor $E$ iteratively extracts witness for each
accepting  session. Specifically, for any $i$, $1\leq i\leq s(n)$,
we denote by $E_i$ the experiment for the knowledge-extractor on
the $i$-th session. $E_i$ emulates $S$ with the \emph{fixed}
random coins included in $sta$, with the exception that the random
coins to be used by the simulator (emulating the honest verifier)
for Stage-3 (i.e., SWIA/POK) of the $i$-th session are no longer
emulated internally, but received externally. 
The experiment $E_i$ amounts to the execution of the SWIA/POK
between  a stand-alone (deterministic) prover and an honest
verifier on common input $(x_i, PK, c^{(i)}_w)$, where $c^{(i)}_w$
is the Stage-2 message sent by $P^*$ in the $i$-th session.
Suppose the $i$-th session w.r.t. common input $x_i$ is accepting
(note that otherwise we do not need to extract a witness and the
witness is set to be ``$\bot$"), by applying  the stand-alone
knowledge-extractor  (for SWIA/POK) on $E_i$, we can extract
$(w_i, r_i)$ in expected polynomial-time.

Here, A subtle point needs to be further clarified. Denote by $p$
the probability that $E_i$ successfully finishes the SWIA/POK on
input $(x_i, c^{(i)}_w)$, by applying the (stand-alone)
knowledge-extractor on $E_i$, we get that the expected
running-time is: $T(n)=p\cdot \frac{q(n)}{p-\kappa(n)}$, where
$\frac{q(n)}{p-\kappa(n)}$ is the running-time of the
knowledge-extractor and $\kappa(\cdot)$ is the knowledge error
function (see Definition \ref{POK}). But, when $p$ is negligible,
as clarified in \cite{L01}, $T(n)$ is not necessarily to be
polynomial in $n$. The technique to deal with this issue is to
apply the technique originally introduced in \cite{GK96} (which is
also deliberated in \cite{L01}). More details about the technique
of dealing with this subtlety are referred to \cite{GK96,L01}.

Now, we consider the value committed to $c^{(i)}_w$ that is also
efficiently extracted. There are three possibilities:

\textbf{Case-1.} $c^{(i)}_w=C(w_i, r_i)$ and  $y_{1-b}=f(w_i)$.
Recall that $PK=(y_0, y_1)$ and $SK=s_b$.

\textbf{Case-2.} $c^{(i)}_w=C(w_i, r_i)$ and $y_{b}=f(w_i)$.

\textbf{Case-3.} $c^{(i)}_{w}=C(w_i, r_i)$ and $(x_i, w_i)\in
R_L$.



Case-1 can occur only with negligible probability, due to the
one-wayness of $f$. Specifically, consider that  $y_{1-b}$ is
given to the simulator as input, rather than being emulated
internally.




Case-2 can occur also with negligible probability, due to the
statistical WI of Stage-1. Suppose Case-2 occurs with
non-negligible probability (and we know Case-1 occurs with
negligible probability), we can simply open $c^{(i)}_w$'s by
brute-force to violate the statistical WI of Stage-1.

By removing Case-1 and Case-2, we conclude now  that for any $i$,
$1\leq i\leq s(n)$, if the $i$-th session in $str$ is accepting
w.r.t. common input $x_i$ selected by $P^*$, then 
$E$ will output a witness $w_i$ for $x_i\in L$.
 To finish the proof,
we need to further show that knowledge-extraction is independent
of the secret-key used by the simulator/extractor (i.e., the joint
KEI property). Specifically, we need to show that $\Pr[R(SK,
\bar{w}, str)=1]$ is negligibly close to $\Pr[R(SK^{\prime},
\bar{w}, str)=1]$ for any polynomial-time computable relation $R$,
where $\bar{w}$ is the list of extracted witnesses (when the
simulator/extractor uses $SK$ as the witness in Stage-1
interactions in $str$) and $SK^{\prime}$ is the  element
(outputted by $S$ in accordance with
$\textsf{Expt}_{\text{CKE}}(1^n, z)$) randomly and
independently distributed over the space of $SK$.  
The joint KEI property is direct from the \emph{statistical} WI of
Stage-1. Specifically, as the extracted witnesses are well-defined
by the statistically-binding $c^{(i)}_w$'s, if the joint KEI
property does not hold,  we directly extract by brute-force   all
witnesses $w_i$'s from $c^{(i)}_w$'s from successful sessions, and
then apply the assumed existing distinguishable relation $R$ to
violate the statistical WI of
Stage-1. 

 In more details, for any pair $(s_0, s_1)$ in key-generation stage
 and for any auxiliary information $z$,
  $\Pr[R(SK, \bar{w}, str)=1] =\frac{1}{2}\Pr[R(s_0, \bar{w}, str)=1|S/E \ \text{uses} \ s_0\
\text{in Stage-1 interactions in } str]+\frac{1}{2}\Pr[R(s_1,
\bar{w},\\ str)=1|S/E\  \text{uses}\  s_1\  \text{in Stage-1
interactions in } str]$, and $\Pr[R(SK^{\prime}, \bar{w}, str)=1]
=\frac{1}{2}\Pr[R(s_0, \bar{w}, str)=1|S/E \ \text{uses} \ s_1\
\text{in Stage-1 interactions in } str]+\frac{1}{2}\Pr[R(s_1,
\bar{w}, str)=1|S/E\ \text{uses}\ s_0\  \text{in Stage-1
interactions}]$. Suppose the  KEI property does not hold, it
implies that there exists a bit $\alpha \in \{0, 1\}$ such that
the difference between $\Pr[R(s_\alpha, \bar{w}, str)=1|S/E \
\text{uses} \ s_0\ \text{in Stage-1 interactions in }str]$ and
$\Pr[R(s_\alpha, \bar{w}, str)=1|S/E \ \text{uses} \ s_1\ \text{in
Stage-1 interactions in }str]$ is non-negligible. Now, we can
incorporate the $(s_\alpha, R)$ into a brute-force algorithm in
order to break   the statistical WI of Stage-1. Further details
are omitted here.  Note that the KEI property holds against any
(not necessarily polynomial-time computable) relation $R$. That
is, the protocol depicted in Figure \ref{CKE-generic} is of
\emph{statistical} CKE. \hfill $\square$

\subsection{On the essential role of Strong WI}

We remark that, with respect to the above generic CZK-CKE
implementation depicted in Figure \ref{CKE-generic}, the  SWI at
Stage-3 plays an essential role for achieving  CZK and CKE
properties simultaneously. In particular, we note that regular WI
is insufficient here. On the one hand, we do not know how to prove
the CZK property in general, when SWI is replaced by a regular WI;
On the other hand, as ZK is itself SWI, one may consider to use a
special ZK (e.g., the FSZK which composes two regular WI
sub-protocols) to replace SWI of Stage-3 such that the special ZK
can share the regular WI of Stage-1 in the public-key model, and
thus we only use regular WIPOK at Stage-3. This  in particular
implies a \emph{round-optimal} (i.e., four-round) implementation
by according round combinations. But, such solution loses the CKE
property and even concurrent soundness in general \emph{in the
public-key model} (see the concrete attack to FSZK in the
public-key model \cite{YZ06}). That is, in the security analysis
of the SWI-based generic CZK-CKE implementation, we will rely on
the argument/proof of knowledge of SWI \emph{in the plain model}
that  is not affected by concurrent composition in the plain
model. If we replace the SWI by a ZK protocol \emph{in the BPK
model}, then we may  require the ZK protocol has already been
CKE-secure, which however is our goal here.



Still, in next section, we consider   more efficient CZK-CKE
implementations based on  regular WI. But the situation with such
solutions  turns out to be much subtler.

\section{Efficient CZK-CKE  in the BPK Model} 
\label{efficient}

In this section, we present the efficient constant-round CZK-CKE
arguments for $\mathcal{NP}$ in the BPK model, and the practical
instantiations. The efficient CZK-CKE protocols rely on some minor
complexity leveraging, in  a novel way, to frustrate potential
concurrent MIM. Along the way, we discuss and clarify the various
subtleties.

Recall that for the generic CZK-CKE implementation presented in
Section \ref{genericCKE},  the strong WI at Stage-3  plays an
essential role for the provable security. But, employing strong WI
complicates the protocol structure, and incurs protocol
inefficiency. It would be desirable to still use regular WI at
Stage-3, for conceptual simple protocol structure as well as for
protocol efficiency. To bypass the subtleties of  SWI for the CZK
proof, we employ a double-commitments technique. Specifically, we
require the prover to produce a \emph{double} of
statistically-binding commitments, $c_w$ and $c_{sk}$, before
starting the second WI sub-protocol, where $c_w$ is supposed to
commit to a valid $\mathcal{NP}$-witness for $x\in L$ and $c_{sk}$
is supposed to commit to the preimage of either  $y_0$ or $y_1$.
Double commitments can bypass, by hybrid arguments,  the
subtleties of
 SWI for the CZK proof. But, the provable CKE property with double
commitments turns out to be much subtler, and we have to employ
(some minimal) complexity leveraging, in a novel way,  to
frustrate potential CMIM adversarial strategies. This renders us
an efficient,  as well as conceptually simple,   CZK-CKE solution,
which can be further high practically instantiated for some
number-theoretic languages.




The generic construction is depicted in Figure \ref{cZKCKE}, page
\pageref{cZKCKE} (as the  construction is for $\mathcal{NP}$ via
$\mathcal{NP}$-reduction,  we do not explicitly describe  the
language-selecting machine $\mathcal{M}$ in the protocol
specification).

\begin{center}
 \begin{figure}[!t]
\begin{tabular} {|c|}
 \hline
\begin{minipage}[t]{6.4in} \small \textbf{Key Generation.}  Let $f: \{0, 1\}^n\rightarrow \{0, 1\}^n$ be any
OWF secure against $2^{n^c}$-time adversaries for some constant
$c$, $0<c<1$,  where $1^n$ is the system security parameter. Each
verifier $V$ selects
 random strings   $s_0$, $s_1$ from $\{0, 1\}^n$, randomly selects a bit $b\leftarrow\{0, 1\}$,
 computes $y_b=f(s_b)$ and sets $y_{1-b}=f(s_{1-b})$. $V$  registers $PK=(y_0, y_1)$ in a public file $F$ as its public-key, and keeps $SK=s_b$
 as its secret-key. Define $R_{KEY}=\{((y_0, y_1), s)|y_0=f(s) \vee y_1=f(s) \}$


    \end{minipage}\\ \hline

\begin{minipage}[t]{6.4in} \small
\begin{description}
\item [Common input.] An element $x\in L\cap\{0, 1\}^{poly(n)}$.
Denote by $R_L$ the corresponding $\mathcal{NP}$-relation for $L$.

\item [P private input.] An $\mathcal{NP}$-witness $w\in \{0,
1\}^{poly(n)}$ for $x\in L$. Here, we assume w.l.o.g. that the
witness for any $x\in L\cap\{0, 1\}^{poly(n)}$ is of the same
length $poly(n)$. 
\end{description}
\end{minipage}\\ \hline

 \begin{minipage}[t]{6.4in} \small
 \textbf{Complexity leveraging.}  The system  parameter is $n$,
 but the statistically-binding commitment $c_{sk}$
 is generated on a relatively smaller security parameter $n_{sk}$. Specifically, suppose the one-wayness of verifier's
 public-key holds against $2^{n^c}$-time adversaries for some constant $c$,  $0<c<1$. Let $\lambda$ be any constant such that
 $\lambda>\frac{1}{c}$, then we set $n=n_{sk}^{\lambda}$. Note
that $n$ and $n_{sk}$ are still polynomially related. That is, any
quantity that is a polynomial of $n$ is also another polynomial of
$n_{sk}$. This complexity leveraging guarantees that although a
$poly(n)\cdot 2^{n_{sk}}$-time adversary can break the hiding
property of $c_{sk}$ on a security parameter $n_{sk}$,  it is
still infeasible to break the one-wayness of $f$ (because
$poly(n)\cdot 2^{n_{sk}}\ll 2^{n^c}$).

    \end{minipage}\\ \hline

\begin{minipage}[t]{6.4in} \small

\begin{description}
\item \item [Stage-1.] $V$ proves to $P$ that it knows a preimage
to  one of $y_0, y_1$, by running a \emph{statistical}  WIA/POK
protocol,  in which $V$ plays the role of knowledge prover. The
witness used by $V$ in this stage is $s_b$.

\item [Stage-2.] If $V$ successfully finishes Stage-1, $P$ does
the following: it computes and sends $c_w=C(w, r_w)$ and
$c_{sk}=C(0^n, r_{sk})$, where $C$ is a statistically-binding
commitment scheme and $r_w$ and $r_{sk}$ are the randomness used
for commitments. $c_{sk}$ is generated on the smaller security
parameter $n_{sk}$ specified above.

\item[Stage-3.]  Define a new $\mathcal{NP}$-language
$L^{\prime}=\{(x, y_0, y_1, c_w, c_{sk})| (\exists (w, r_w)\  s.t.
\ c_w= C(w, r_w) \wedge (x, w)\in R_L) \vee (\exists (w,
r_{sk},b)\ s.t.\   c_{sk}=C(w, r_{sk}) \wedge y_b=f(w) \wedge b\in
\{0, 1\})\}$. Then,   $P$ proves to $V$ that it knows a witness
for $(x, y_0, y_1, c_w, c_{sk})\in L^{\prime}$, by running a
(3-round)  WI argument/proof of knowledge (WIA/POK) protocol for
$\mathcal{NP}$
 (e.g., the $n$-parallel repetition of Blum's protocol for DHC). 

\end{description}
\end{minipage}


\\
\hline
\end{tabular}

\caption{\label{cZKCKE}
  The efficient CZK-CKE argument $\langle P, V \rangle$ for $\mathcal{NP}$ in the BPK
  model}
\end{figure}
\end{center}

\textbf{Note on efficiency.} Though we employ double
 commitments at Stage-2,  the strong WIA/POK of
Stage-3 in the generic construction is replaced by any regular
WIA/POK here, from which we can  gain much better  efficiency
advantage. In particular, as we shall see, the efficient
construction can be high practically instantiated. It's also easy
to see that the implementation  can be round-optimal by round
combinations.

\textbf{Notes on the complexity leveraging.} We remark  that
complexity leveraging via the sub-exponential hardness assumption
on verifier's public-key is only for provable  security analysis
to frustrate concurrent MIM. Both CZK simulation and CKE
knowledge-extraction are still polynomial-time. We note that the
use of complexity leveraging for frustrating concurrent MIM could
be a novel paradigm,  different from the uses of complexity
leveraging in existing works for protocols in the BPK model (e.g.,
\cite{CGGM00}). Such paradigm can also be applied to other
scenarios  to frustrate potential concurrent MIM, while still
providing polynomial-time simulation and/or knowledge-extraction.
  Note  also
that the complexity leveraging is minimal: it only applies to
$c_{sk}$ and all other components of the protocol work on the
general system parameter $n$;  also,  all components except for
verifier's public-keys can be standard polynomially secure.
Furthermore, as we shall see,  the complexity leveraging can be
waived as long as only concurrent soundness is concerned. 
    We remark that though non-standard, sub-exponential hardness assumption may still be
viewed to be reasonable, which is also used in a large body of
works for fulfilling various cryptographic tasks. Detailed
discussions and clarifications of the use of complexity leveraging
for frustrating concurrent MIM
can be found in 
 Section \ref{insufficiency}.

\textbf{On the  necessity of \emph{double} commitments $c_w$ and
$c_{sk}$.}  We stress that in the context of the above protocol
structure of efficient CZK-CKE, mandating \emph{double}
commitments $c_w$ and $c_{sk}$ of Stage-2 plays a very crucial
role for \emph{simultaneously} achieving CZK and CKE in the
public-key model. On the one hand, for protocol variants without
either $c_w$ or $c_{sk}$, concrete attacks exist, showing that
they are not concurrently knowledge-extractable. 
 Details
are presented in Section \ref{necessity}; On the other hand,
double commitments enable  us to bypass the need of \emph{strong}
WI of Stage-3
 for correct CZK simulation. Specifically, by
employing double commitments the CZK simulation is not based on
the strong WI property of Stage-3, and it is shown that  regular
WI is sufficient  for correct CZK simulation by  hybrid arguments.

\subsection{Security analysis}\label{analysis}

\textbf{Notes on the underlying hardness assumptions and
round-complexity.} First note that except for subexponential
hardness assumption on the OWF $f$ used in key generation, all
other components in our solution can be standard polynomially
secure. We note that if the OWF $f$  admits perfect/statistical
$\Sigma$-protocols (and thus we can use $\Sigma_{OR}$ in Stage-1),
the protocol depicted in Figure \ref{cZKCKE} can be based on any
 sub-exponentially strong OWF admitting perfect/statistical
$\Sigma$-protocols, and be of optimal (i.e., 4-round)
round-complexity by  round combinations; If we use in Stage-1 the
modified Blum's protocol for DHC with constant-round
statistically/perfectly hiding commitments, the protocol depicted
in Figure \ref{cZKCKE} can be based on any collision-resistant
hash function and any sub-exponentially strong OWF with optimal
round-complexity, or based on any sub-exponentially strong
claw-free collection (with efficiently recognizable index set) but
with 5 rounds. In the later case (with modified Blum's protocol
for DHC), we can use any sub-exponentially strong OWF for key
generation.


\begin{theorem}
The protocol depicted in Figure \ref{cZKCKE} is   concurrently
knowledge-extractable concurrent ZK argument for $\mathcal{NP}$ in
the BPK  model.
\end{theorem}

\textbf{Proof (sketch).} The completeness of the protocol $\langle
P, V\rangle$ can be easily checked.

\textbf{Concurrent zero-knowledge.}

We first consider a mental simulator $M$ that takes as input all
secret-keys corresponding to all public-keys registered in the
public-key file, \emph{in case the corresponding secret-keys
exist}.

For any $s(n)$-concurrent malicious verifier $V^*$ (defined in
Section \ref{APPBPK}) and any $\mathcal{NP}$-language $L$, $M$
runs $V^*$ as a subroutine on inputs $\bar{\textbf{x}}=\{x_1,
\cdots, x_{s(n)}\}\in L^{s(n)}$ (where $x_i$ might equal $x_j$,
$1\leq i, j \leq s(n)$ and $i\neq j$), the public file $F=\{PK_1,
\cdots, PK_{s(n)}\}$ and all assumed existing secret-keys. $M$
works just as the honest prover does in Stage-1 of any session. In
Stage-2 of any session on a common input $x_i$ and with respect to
a public-key $PK_j$ (i.e., the $i$-th session w.r.t $PK_j$, $1\leq
i, j \leq s(n)$),  $M$ computes $c^{(i)}_w=C(0^{poly(n)},
r^{(i)}_w)$ and $c^{(i)}_{sk}=C(SK_j, r^{(i)}_{sk})$, where $SK_j$
is the secret-key corresponding to $PK_j$  for which we assume it
exists and $M$ knows. Then, $M$ runs the WIA/POK protocol with
$V^*$ in Stage-3 of the session with $(SK_j, r^{(i)}_{sk})$ as its
witness.

To show the output of $M$ is indistinguishable from the view of
$V^*$ in real concurrent interactions, we consider another mental
simulator $M^{\prime}$. $M^{\prime}$ takes both the witnesses for
$\bar{\textbf{x}}=\{x_1, \cdots, x_{s(n)}\}$ and all the
secret-keys corresponding to public-keys registered in $F$ (in
case the corresponding secret-keys exist). $M^{\prime}$ works just
as $M$ does, but with the following exception: for any $i$, $j$,
$1\leq i, j \leq s(n)$, in Stage-2 of the $i$-th session on common
input $x_i$ w.r.t $PK_j$,  $M^{\prime}$ computes $c^{(i)}_w=C(w_i,
r^{(i)}_w)$, where $w_i$ is the witness for the common input
$x_i$. Note that the witness used by $M^{\prime}$ in Stage-3 is
still $SK_j$, just as $M$ does. That the output of $M^{\prime}$ is
indistinguishable from that of $M$ is from the computational
hiding property of the statistically-binding commitment scheme $C$
used in Stage-2. Otherwise, by a simple hybrid argument, we can
violate the hiding property of the underlying
 commitment scheme $C$.

 We now consider another mental simulator $M^{\prime\prime}$ that
 mimics $M^{\prime}$  with the following exception: for any $i$, $j$, $1\leq i,
j \leq s(n)$, in Stage-3 of  the $i$-th session on common input
$x_i$ w.r.t $PK_j$, the witness used by $M^{\prime\prime}$ is
$w_i$, rather than $SK_j$ as used by $M^{\prime}$. By hybrid
arguments, the output of $M^{\prime\prime}$ is indistinguishable
from that of $M^{\prime}$ by the WI property 
  of Stage-3. Also, by hybrid arguments, the
output of $M^{\prime\prime}$ is also indistinguishable from the
view of $V^*$ in real concurrent interactions by the computational
hiding property of the underlying  commitment scheme $C$ used in
Stage-2.

This establishes  that the output of $M$ is indistinguishable from
the view of $V^*$ in real concurrent interactions. To build a PPT
simulator $S$  from scratch, where $S$ does not know any
secret-keys corresponding to public-keys in the public file, we
again  resort to the
 technique developed in \cite{CGGM00}.
Specifically, $S$ works in $s(n)+1$ phases. In each phase, $S$
either successfully finishes the simulation, or ``covers"  a  new
public-key  for which it has not known the corresponding
secret-key up to now  in case $V^*$ successfully finishes the
Stage-1 interactions w.r.t. that public-key.  Key covering is
guaranteed by the POK property of Stage-1 interactions. For more
details, see \cite{CGGM00}.

\textbf{(Statistical) concurrent knowledge-extraction.}

According to the CKE formulation, for any $s$-concurrent malicious
prover $P^*$ (defined in Section \ref{basic}) we need to build two
algorithms $(S, E)$. The simulator $S$, on inputs $(1^n, z)$,
works as follows: It first perfectly emulates the key-generation
stage of the honest verifier, getting $PK=(y_0, y_1)$ and $SK=s_b$
and $SK^{\prime}=s_{1-b}$ for a random bit $b$. Then, $S$ runs
$P^*$ on $(1^n, PK, z)$ to get $(R_L, \tau)$, where $R_L$
indicates an $\mathcal{NP}$-language for which the proof-stages
will work and $\tau$ is some auxiliary information to be used by
$P^*$ in proof-stages. In the proof stages, $S$ perfectly emulates
the honest verifier with the secret-key $SK$. Finally, whenever
$P^*$ stops, $S$ outputs the simulated transcript $str$, together
with the state information $sta$ set to be $(PK, SK, SK^{\prime},
z)$ and the random coins used by $S$.  Note that the simulated
transcript $str$ is identical to the view  of $P^*$ in real
execution.

The knowledge-extraction process is similar to that of
\cite{PR05f}. Note that we need to extract witnesses to all
accepting sessions in $str$.  Given $(str, sta)$, the
knowledge-extractor $E$ iteratively extracts witness for each
accepting  session. Specifically, for any $i$, $1\leq i\leq s(n)$,
we denote by $E_i$ the experiment for the knowledge-extractor on
the $i$-th session. $E_i$ emulates $S$ with the \emph{fixed}
random coins included in $sta$, with the exception that the random
challenge (i.e., the second-round message) of the WIA/POK protocol
of Stage-3 in the $i$-th session is no longer
emulated internally, but received externally. 
The experiment $E_i$ amounts to the execution of the WIA/POK
protocol  of Stage-3 between a stand-alone (deterministic) prover
and an honest verifier on common input $x_i$. Suppose the $i$-th
session w.r.t. common input $x_i$ is accepting (note that
otherwise we do not need to extract a witness and the witness is
set to be ``$\bot$"), by applying  the stand-alone
knowledge-extractor (for the underlying WIA/POK)  on $E_i$,
according to the POK property of the underlying WIA/POK protocol
(say, the $n$-parallel repetition of Blum's protocol for DHC)
 except
for the probability $2^{-n}$ we can extract $(w_i, r_i)$ in
expected polynomial-time, satisfying one of the following:

\textbf{Case-1.} $c^{(i)}_{sk}=C(w_i, r_i)$ and  $y_{1-b}=f(w_i)$,
where $c^{(i)}_{sk}$ and $c^{(i)}_w$ are the double
statistically-binding commitments sent at the Stage-2 of the
$i$-th session, and $SK=s_b$.

\textbf{Case-2.} $c^{(i)}_{sk}=C(w_i, r_i)$ and $y_{b}=f(w_i)$.

\textbf{Case-3.} $c^{(i)}_{w}=C(w_i, r_i)$ and $(x_i, w_i)\in
R_L$.



Case-1 can occur only with negligible probability, due to the
one-wayness of $f$. Specifically, consider that  $y_{1-b}$ is
given to the simulator as input, rather than being emulated
internally.

The subtle point here is: by applying the stand-alone
knowledge-extractor on $E_i$, the Stage-1 interactions given by
the simulator/extractor would also be rewound, which could reveal
the secret-key $SK$. In particular, recall  the  adversarial
strategies presented  in Section \ref{appattack}. Here, it is the
critical combination of complexity leveraging on the
statistically-binding commitment $c_{sk}$ and the statistical WI
of Stage-1 that provably rules out such concurrent interleaving
and malleating attacks.


\begin{proposition} \label{pcase2}
Case-2 occurs with negligible probability. \end{proposition}

\textbf{Proof} (of Proposition \ref{pcase2}). Suppose Case-2
occurs
with non-negligible probability, this means that 
  for some $(s_0,
s_1, b)$, where $s_0, s_1 \in \{0, 1\}^n$ and  $b\in \{0, 1\}$,
such that when the simulator $S$ uses $s_b$ as the witness for
simulating Stage-1 interactions, with non-negligible probability
$p(n)$,  the $c^{(i)}_{sk}$ in the simulated transcript $str$
outputted by $S$ is a commitment of $s_b$. Otherwise, Case-2 will
trivially occur with negligible probability.  But, due to the
statistical WI of Stage-1, with the same probability $p(n)$ the
$c^{(i)}_{sk}$ in the simulated transcript $str$ outputted by $S$,
when it uses $s_{1-b}$ as the witness for simulating Stage-1
interactions, is still a commitment of $s_b$. Note that the  value
committed in $c^{(i)}_{sk}$ can be brute-force extracted in time
$poly(n)\cdot 2^{n_{sk}}\ll 2^{n^c}$. Now, suppose $y_b=f(s_b)$ is
given to the simulator as input externally, and $y_{1-b}$ and
Stage-1 interactions are simulated by the simulator (with
$s_{1-b}$ as the witness), this implies that there exists  an
algorithm  that can break the one-wayness of $y_b$ in
$poly(n)\cdot 2^{n_{sk}}\ll 2^{n^c}$-time, which violates the
sub-exponential hardness of $y_b$.

\textbf{On the subtleties without  the complexity leveraging}. We
remark that  the uses of the  complexity leveraging on $c_{sk}$,
along with  statistical WI of Stage-1,  not only provably rules
out Case-2, but also \emph{greatly simplifies} the proof of
Proposition \ref{pcase2}. In particular,  we do not know how to
provably prove Proposition \ref{pcase2} without the complexity
leveraging.
 Detailed clarifications of the subtleties   are presented
in Section \ref{insufficiency}, which in particular implies that
the efficient CZK-CKE protocol depicted in Figure \ref{cZKCKE} is
concurrently sound \emph{under standard polynomial-time hardness
assumptions}. 
 \hfill $\square$


By removing Case-1 and Case-2, we conclude now  that for any $i$,
$1\leq i\leq s(n)$, if the $i$-th session in $str$ is accepting
w.r.t. common input $x_i$ selected by $P^*$, then 
$E$ will output a witness $w_i$ for $x_i\in L$.
 To finish the proof,
we need to further show that knowledge-extraction is independent
of the secret-key used by the simulator/extractor (i.e., the joint
KEI property). Specifically, we need to show that $\Pr[R(SK,
\bar{w}, str)=1]$ is negligibly close to $\Pr[R(SK^{\prime},
\bar{w}, str)=1]$ for any polynomial-time computable relation $R$,
where $\bar{w}$ is the list of extracted witnesses (when the
simulator/extractor uses $SK$ as the witness in Stage-1
interactions in $str$) and $SK^{\prime}$ is the  element
(outputted by $S$ in accordance with
$\textsf{Expt}_{\text{CKE}}(1^n, z)$) randomly and
independently distributed over the space of $SK$.  
The joint KEI property is direct from the \emph{statistical} WI of
Stage-1. Specifically, as the extracted witnesses are well-defined
by the statistically-binding $c^{(i)}_w$'s, if the joint KEI
property does not hold,  we directly extract by brute-force   all
witnesses $w_i$'s from $c^{(i)}_w$'s of successful sessions, and
then apply the assumed existing distinguishable relation $R$ to
violate the statistical WI of
Stage-1. 

 In more details, for any pair $(s_0, s_1)$ in key-generation stage
 and for any auxiliary information $z$,
  $\Pr[R(SK, \bar{w}, str)=1] =\frac{1}{2}\Pr[R(s_0, \bar{w}, str)=1|S/E \ \text{uses} \ s_0\
\text{in Stage-1 interactions in } str]+\frac{1}{2}\Pr[R(s_1,
\bar{w},\\ str)=1|S/E\  \text{uses}\  s_1\  \text{in Stage-1
interactions in } str]$, and $\Pr[R(SK^{\prime}, \bar{w}, str)=1]
=\frac{1}{2}\Pr[R(s_0, \bar{w}, str)=1|S/E \ \text{uses} \ s_1\
\text{in Stage-1 interactions in } str]+\frac{1}{2}\Pr[R(s_1,
\bar{w}, str)=1|S/E\ \text{uses}\ s_0\  \text{in Stage-1
interactions}]$. Suppose the  KEI property does not hold, it
implies that there exists a bit $\alpha \in \{0, 1\}$ such that
the difference between $\Pr[R(s_\alpha, \bar{w}, str)=1|S/E \
\text{uses} \ s_0\ \text{in Stage-1 interactions in }str]$ and
$\Pr[R(s_\alpha, \bar{w}, str)=1|S/E \ \text{uses} \ s_1\ \text{in
Stage-1 interactions in }str]$ is non-negligible. Now, we can
incorporate the $(s_\alpha, R)$ into a brute-force algorithm in
order to break   the statistical WI of Stage-1. Further details
are omitted here.  
 Note that the KEI property holds against any (not necessarily
polynomial-time computable) relation $R$, that is, the protocol
depicted in Figure \ref{cZKCKE} is of \emph{statistical} CKE.
\hfill $\square$

\subsection{On the subtleties without the complexity
leveraging} \label{appmimdiscussion} \label{insufficiency}

In this section, we clarify the subtleties and justify the
necessity of the (minimal) complexity leveraging  on $c_{sk}$ with
  the efficient CZK-CKE. We first give high-level discussions on the use of
complexity leveraging against (concurrent) men-in-the-middle;
Then, we  make in-depth clarifications  by attempting  to provide
a proof of Proposition \ref{pcase2} without the complexity
leveraging on $c_{sk}$, which  identifies the subtleties  or
difficulties that seemingly cannot be overcome without exploiting
 the complexity leveraging on $c_{sk}$ (and also the statistical
 WI of Stage-1).


\subsubsection{On the use of complexity leveraging against
man-in-the-middle}\label{highlevelMIM}
 Recall that, for  the generic CZK-CKE (depicted in Figure \ref{CKE-generic}),
 to successfully finish the $i$-th session with commit-then-SWI mechanism,
for any $i$, $1\leq i\leq s(n)$,  an $s$-concurrent  adversary
$P^*$  has to use the value committed to (determined by) the
\emph{unique} Stage-2 commitment $c^{(i)}_w$ as the witness in
Stage-3 SWI. But, for the efficient CZK-CKE, $P^*$ however has
\emph{double} choices: it can use either the value committed to
$c^{(i)}_{sk}$ or the value committed to $c^{(i)}_w$, as the
witness in  Stage-3 regular WI. 
 We consider two potential adversarial strategies:

\begin{description}
\item [Adversarial-Strategy-1.] $P^*$ commits a valid witness $w$
(for $x_i \in L$) to $c^{(i)}_w$,  and commits  a secret-key, say
$s_0$, to $c^{(i)}_{sk}$ in Stage-2 of the $i$-th session
(\emph{possibly by malleating verifier's public-keys into $x_i$
and $c^{(i)}_{sk}$}), where $x_i$ is the common input adaptively
selected by $P^*$ for the $i$-th session; Then, \emph{possibly by
malleating the Stage-1 concurrent interactions}, $P^*$ always uses
the valid witness $w$ in Stage-3 of the $i$-th session  in case
the honest verifier $V$ uses $s_1$ as the witness in Stage-1
interactions (\emph{note that $w$ could be maliciously related to
$s_1$ as well,  as the common input $x_i$ is selected by $P^*$}),
but uses $s_0$ as the witness in Stage-3 with non-negligible
probability in case  $V$ uses $s_0$ as the witness in Stage-1
interactions.


\item [Adversarial-Strategy-2.]  With non-negligible probability
$p$, $P^*$ commits $s_0$ (resp., $s_1$) to $c^{(i)}_{sk}$ in
Stage-2 of the $i$-th session (again, \emph{possibly by malleating
verifier's public-keys into $c^{(i)}_{sk}$}); Then, \emph{possibly
by malleating the Stage-1 concurrent interactions}, $P^*$
successfully finishes  Stage-3
 of the session with $s_0$ (resp., $s_1$) as the
witness, in case $V$ uses $s_0$ (resp., $s_1$) as the witness in
Stage-1 interactions; However, with the same probability $p$,
$P^*$ commits both a valid witness $w$ to $c^{(i)}_w$ and $s_0$
(resp. $s_1$) to $c^{(i)}_{sk}$ in Stage-2 of the session,  and
successfully finishes  Stage-3  with $w$ as the witness in case
$V$ uses $s_1$ (resp., $s_0$) as the witness in Stage-1
interactions.


\end{description}

Note that the concurrent malicious prover $P^*$
 actually amounts to a concurrent MIM who manages,
by concurrent interleaving interactions, to malleate verifier's
public-keys and Stage-1 interactions (in which it plays the role
of the verifier) into successful Stage-2 and Stage-3 interactions
(in which $P^*$  plays the role of the prover),  but without
knowing
any witness for the Stage-2 and Stage-3 interactions. 
 Note that both the above two cases indicate the failure
of knowledge-extraction correctness: that is, with non-negligible
probability, the value extracted (when using $SK=s_b$ \emph{for a
random bit $b$}) is the preimage of $y_0$ or $y_1$ committed to
$c^{(i)}_{sk}$. But, no contradiction can be reached without
resorting to the complexity leveraging. In particular, they do not
violate the statistical WI of Stage-1: in the first case, the
value committed to $c^{(i)}_{sk}$ is fixed; and in the second
case, with probability $2p$, the value committed to $c^{(i)}_{sk}$
is $s_b$ for \emph{both} $b\in \{0, 1\}$, no matter which
secret-key (whether $s_0$ or $s_1$) is used in Stage-1
interactions.   As we do not employ any non-malleable building
tools and we  are actually facing a concurrent MIM $P^*$, the
above MIM  adversarial strategies could indeed be potential. At
least, we do not know how to \emph{provably} rule out such
seemingly impossible adversarial activities, without resorting to
the complexity leveraging.



  We note that the use of complexity leveraging
for frustrating concurrent MIM could be a novel paradigm,
different from the uses of complexity leveraging in existing works
  (e.g., \cite{CGGM00,YZ07}). 
 Such paradigm may  be possibly of independent interest, and can   be
applied in other scenarios to frustrate potential concurrent MIM,
while still providing polynomial-time simulation and/or
knowledge-extraction \emph{as well as remaining the protocol
efficiency and conceptual simple  protocol  structure}. Note also
that the complexity leveraging is minimal:
 it only applies to $c_{sk}$,  
 and all components except for verifier's public-keys can be standard
polynomially secure. 

\subsubsection{Analysis attempt without complexity leveraging}

In this section, by attempting  to provide a proof of Proposition
\ref{pcase2} without the complexity leveraging on $c_{sk}$,  we
clarify the subtleties or difficulties that seemingly cannot be
overcome without exploiting
 the complexity leveraging on $c_{sk}$  (and also the statistical WI of
Stage-1). The analysis in particular implies that  the efficient
CZK-CKE protocol depicted in Figure \ref{cZKCKE} is concurrently
sound \emph{under standard polynomial-time hardness assumptions}
and that \emph{partial witness independent} WI (employed in the
works of \cite{DV05,DL06,DDL06}) seems  to be insufficient  even
for correct knowledge-extraction for \emph{individual} statements.
In the following security analysis, we assume no complexity
leveraging on $c_{sk}$, i.e., verifier's public-keys are standard
polynomially secure and $c_{sk}$ is formed on the same system
parameter $n$.

  We consider two experiments:
$\mathcal{E}_0$ and $\mathcal{E}_1$. For each $\mu \in \{0, 1\}$,
 $\mathcal{E}_\mu$ mimics the experiment $E_i$ (specified in the security analysis in
 Section  \ref{analysis}),
with the following exceptions:  $\mathcal{E}_\mu$  uses $s_\mu$ as
its witness in Stage-1 interactions  (note that $(s_0, s_1)$ is
included in $sta$);  and the   coins used by $\mathcal{E}_{\mu}$
for internal emulation of the \emph{proof stages} are randomly and
independently chosen (i.e., they are independent of the coins
included in $sta$); The coins for the first-stages of $V$ and
$P^*$   are still  those fixed in $sta$, with respect to which we
suppose Case-2 will occur with non-negligible probability.
 Suppose Case-2 occurs with non-negligible probability, then there must exist
a bit $\mu$ such that applying the (stand-alone)
knowledge-extractor on  $\mathcal{E}_\mu$ will output the preimage
of $y_\mu$ with non-negligible probability. Otherwise, Case-2 will
trivially occur with negligible probability. Without loss of
generality, we assume $\mu=0$. That is,  the knowledge-extractor
on $\mathcal{E}_0$ outputs the preimage of $y_0$ with
non-negligible probability (and outputs the preimage of $y_1$ with
negligible probability due to the one-wayness of $f$). Now we
consider the output of the knowledge-extractor on $\mathcal{E}_1$:
first, it outputs the preimage of $y_0$ also with negligible
probability; thus, with non-negligible probability (as we assume
Case-2 occurs with non-negligible probability and Stage-1
interactions are WI),  the knowledge-extractor on $\mathcal{E}_1$
outputs either the preimage of $y_1$ or the witness for some $x\in
L$ where $x$ is the common input of the $i$-th session in
$\mathcal{E}_1$. Note that $x$ is not necessarily the same $x_i$
in $E_i$ as the coins used by $\mathcal{E}_{\mu}$ are not the same
as  those of $E_i$.

\textbf{Note.} Here, we cannot directly conclude that the
knowledge-extractor on $\mathcal{E}_1$ will certainly output the
preimage of $y_1$ with non-negligible probability, as we cannot
rely
on the assumption that $x\not\in L$.  
This point complicates the security analysis, and is one
underlying reason for requiring  the complexity leveraging.

Now,  we want to contradict the statistical WI property of
Stage-1.
 We define a series of hybrid mental experiments $H_1,
\cdots, H_{s(n)}$ as follows: for any $k$, $1\leq k \leq s(n)$,
$H_k$ mimics the behavior of $\mathcal{E}_0$ but with the
following exceptions: In Stage-1 of the first $k$ sessions $H_k$
uses $s_1$ as its witness; and in Stage-1 of the rest $s(n)-k$
sessions it uses $s_0$ as the witness. Note that $H_0$ equals the
experiment $\mathcal{E}_0$, and $H_{s(n)}$ equals the experiment
$\mathcal{E}_1$. As we assume that the (stand-alone)
knowledge-extractor on $H_0 (=\mathcal{E}_0$) will output the
preimage of $y_0$ with non-negligible probability (but output the
preimage of $y_1$ with negligible probability), and that the
knowledge-extractor on  $H_{s(n)} (=\mathcal{E}_1)$ will output
either a preimage of $y_1$ or a witness for some $x\in L$ with
non-negligible probability (but output the preimage of $y_0$ only
with negligible probability). By hybrid arguments, we conclude
that there must exist a $k$, $1\leq k\leq s(n)$, such that the
knowledge-extractor on $H_{k-1}$ outputs the preimage of $y_0$
with non-negligible probability and the knowledge-extractor on
$H_k$ outputs the preimage of $y_0$ with negligible probability
(and outputs the preimage of $y_1$ or a witness for some $x\in L$
with non-negligible probability). \emph{Recall that, in all the
experiments, the (stand-alone) knowledge-extractor is to extract
the knowledge for the statement whose validity was successfully
conveyed  in the $i$-th session.} Then we attempt to break the
statistical WI property or Stage-1, by considering another
experiment $B$.

$B$ mimics  $H_k$ with the following exceptions: The Stage-1
interactions of the $k$-th session are no longer emulated
internally, but interacting externally with an external
knowledge-prover  $\hat{P}_k$ who uses $s_\delta$ as the witness
for a random bit $\delta$.
Note that, if $\hat{P}_k$ uses $s_1$ as its witness then the
experiment $B$ is identical to $H_k$, and   if $\hat{P}_k$ uses
$s_0$ as its witness then $B$ is identical to $H_{k-1}$. Now, we
consider two cases:

\begin{description}
\item [Case-2.1.] The external interactions with $\hat{P}_k$ have
finished before the sending of the random challenge (i.e., the
second-round message)  of Stage-3 of the $i$-th session.

\item [Case-2.2.] The external interactions with $\hat{P}_k$ have
not finished on the sending of the  random challenge of Stage-3 of
the
$i$-th session. 
 Note that  the concurrent interleaving and
malleating attack described in Section \ref{attack} is just a
demonstration  of this case.
\end{description}

If Case-2.1 occurs,  we break the WI property of Stage-1 as
follows: Note that in this case, applying the stand-alone
knowledge-extractor on (the $i$-th session in)  $B$ does not incur
rewinding the interactions with $\hat{P}_k$. We can combine the
stand-alone knowledge-extractor and the internal emulation of $B$
into a stand-alone (expected polynomial-time) knowledge-verifier
interacting with $\hat{P}_k$. If the knowledge-extractor  outputs
the preimage of $y_0$, then we also output 0; in any other case,
we output a random bit. According to the above hybrid arguments,
if $\hat{P}_k$ uses $s_0$ as its witness, then we will output 0
with probability that is non-negligibly bigger than 1/2; on the
other hand, if $\hat{P}_k$ uses $s_1$ as its witness, then we will
output 0 with probability negligibly close to 1/2. Furthermore,
using Markov's  inequality,  standard technique (as is done in
\cite{P03C,YZ05}) shows that: if the WI property holds w.r.t. any
strict polynomial-time algorithm it also holds with any expected
polynomial-time algorithm. This contradicts the WI property of the
underlying protocol. Note that \emph{computational} WI of Stage-1
is sufficient for ruling out Case-2.1.

If Case-2.2 occurs, 
 we further distinguish two cases according to the output of the  knowledge-extractor
 on $H_k$. 
Recall that we have assumed that the output of the
knowledge-extractor on $H_k$ is the preimage of $y_0$
 only with negligible probability, and   the output
 of the stand-alone knowledge-extractor
 on $H_{k-1}$ is the preimage of $y_0$ with non-negligible probability.

 \begin{description}

\item [Case-2.2.1.] With \emph{negligible} probability the output
of the (stand-alone) knowledge-extractor on $H_k$ is $s_1$ (i.e.,
the output is always  a witness for some $x\in L$ of the $i$-th
session in $H_k$). 
 This case can be partially illustrated by the Adversarial-Strategy-1 demonstrated in
Section \ref{highlevelMIM}.


\textbf{Note.} It is easy to see that, suppose the common input
$x$ of the $i$-th session in $H_k$ is false, i.e., $x\not \in L$,
then Case-2.2.1 
 can appear at most with
negligible probability.
 We note that \emph{partial witness independent} WI (employed in the
works of \cite{DV05,DL06,DDL06}) seems  to be insufficient even
for correct knowledge-extraction for \emph{individual} statements
(recall that our CKE formulation is w.r.t. \emph{joint}
knowledge-extraction for all statements whose validity was
successfully conveyed in the concurrent sessions). This point was
not addressed in existing works. In particular, with respect to
the Adversarial-Strategy-1, in this case the knowledge-extractor
will extract a secret-key $s_0$ with non-negligible probability
when it simulates Stage-1 interactions with $s_0$ as the witness,
which indicates the failure of correct knowledge-extraction
\emph{even for any individual statement}. 

 \item [Case-2.2.2.]
 With non-negligible probability the output of the stand-alone knowledge-extractor on $H_k$ is the preimage of
 $y_1$.  This case can be partially  illustrated by
 Adversarial-Strategy-2 demonstrated in Section
 \ref{highlevelMIM}.


\textbf{Note.}  Again,  suppose the common input  $x$ of the
$i$-th session in $H_k$ is false, i.e., $x\not \in L$, then
Case-2.2.2 can appear at most with negligible probability.
Otherwise, the value committed in $c^{(i)}_{sk}$ indicates the
secret-key  used in Stage-1 interactions. Recall that we have
assumed that the output of the knowledge-extractor on $H_k$ is the
preimage of $y_0$
 only with negligible probability, and   the output
 of the stand-alone knowledge-extractor
 on $H_{k-1}$ is the preimage of $y_0$ with non-negligible
 probability.  Specifically, suppose the
witness used for Stage-1 interactions is $s_b$, then the
\emph{successful} $i$-th session with $c^{(i)}_{sk}$ committing to
$s_{1-b}$ occurs with negligible probability (\emph{conditioned on
$x\not\in L$}). This  violates the \emph{statistical} WI of
Stage-1.

\end{description}

\textbf{Remark.} Although it intuitively seems that Case-2.2 (in
particular,  the   exemplifying adversarial strategies) could not
occur with non-negligible probability, it (and particularly  the
exemplifying  adversarial strategies presented in Section
\ref{highlevelMIM}) could indeed be potential, as we do not employ
any non-malleable building tools and we are actually facing a
concurrent MIM. We do
not know how to \emph{provably} rule out such possibilities, 
 without resorting to the complexity leveraging  on $c_{sk}$.

\subsection{On the necessity of double commitments}\label{necessity} To show the necessity of the
double commitments $c_w$ and $c_{sk}$ used in Stage-2 of the
efficient CZK-CKE protocol depicted  in Figure \ref{cZKCKE}, we
demonstrate concrete attacks against variants of the protocol
without either $c_w$ or $c_{sk}$, where  
  WIA/POK  protocols are implemented by $\Sigma_{OR}$-protocols. 

\subsubsection{The attack against variant protocol without $c_{w}$}
The variant protocol without $c_{w}$, which amounts to the CZK
protocols of \cite{Z03,DL06}, 
 is re-depicted in Figure \ref{withoutcw} (page
\pageref{withoutcw}).

\textbf{On the implementations of $\Sigma_{OR}$.} For the
$\Sigma_{OR}$-based protocol variant depicted in Figure
\ref{withoutcw}, to get \emph{statistical} WI of Stage-1 there are
two ways: In particular, we can require the underlying OWF $f$
used in the key-generation stage admits perfect/statistical
$\Sigma$-protocols, and thus the $\Sigma_{OR}$ of Stage-1 is
perfect/statistical WI; In general, the variant of (the
$n$-parallel repetition of)  Blum's protocol for DHC, where the
statistically-binding commitments used in the first round are
replaced by the one-round statistically-hiding commitments based
on collision-resistant hash functions, is a \emph{statistical}
$\Sigma$-protocol (as well as statistical WI argument)  for
$\mathcal{NP}$, and thus can be applied to any $\mathcal{NP}$
language under the assumption of collision-resistant hash
functions.

\begin{center}
 \begin{figure}[!t]
\begin{tabular} {|c|}
 \hline \label{withoutcw}
\rule[-3mm]{0mm}{8mm}  \textbf{ $\Sigma_{OR}$-based  protocol
variant without $c_{w}$ \quad $\langle P, V \rangle$}\\ \hline
\begin{minipage}[t]{6.4in} \small \textbf{Key Generation.}
 Let $f: \{0, 1\}^n\rightarrow \{0, 1\}^n$ be any
OWF 
 where
$n$ is the security parameter. Each verifier $V$ selects
 random strings   $s_0$, $s_1$ from $\{0, 1\}^n$, randomly selects a bit $b\leftarrow\{0, 1\}$,
 computes $y_b=f(s_b)$ and sets $y_{1-b}=f(s_{1-b})$. $V$  registers $PK=(y_0, y_1)$ in a public file $F$ as its public-key, and keeps $SK=s_b$
 as its secret-key.
    \end{minipage}\\ \hline

\begin{minipage}[t]{6.4in}
\begin{description}
\item [Common input.] An element $x\in L\cap\{0, 1\}^{poly(n)}$.
Denote by $R_L$ the corresponding $\mathcal{NP}$-relation for $L$.
\item [P private input.] An $\mathcal{NP}$-witness $w\in \{0,
1\}^{poly(n)}$ for $x\in L$.
\end{description}
\end{minipage}\\ \hline

\begin{minipage}[t]{6.4in}

\begin{description}
\item \item [Stage-1.] $V$ proves to $P$ that it knows the
preimage of either  $y_0$ or $y_1$, by running a
$\Sigma_{OR}$-protocol 
 on the input
$(y_0, y_1)$ 
in which $V$ plays the role of the knowledge prover.   The witness
used by $V$ in this stage is $s_b$. Denote by $a_V, e_V, z_V$, the
first-round, the second-round and the third-round message of the
$\Sigma_{OR}$-protocol, respectively. 

\item [Stage-2.] If $V$ successfully finishes Stage-1, $P$ does
the following: it computes $c_{sk}=C(0^n, r_{sk})$, where $C$ is a
perfectly-binding commitment scheme and $r_{sk}$ is  the
randomness used for commitments.

\item[Stage-3.]  Define a new $\mathcal{NP}$-language
$L^{\prime}=\{(x, y_0, y_1, c_{sk})| (\exists w\  s.t. \ (x, w)\in
R_L) \vee (\exists (w, r_{sk}, b)\ s.t.\  c_{sk}=C(w, r_{sk})
\wedge y_b=f(w) \wedge b\in \{0, 1\})\}$. Then, $P$ proves to $V$
that it knows a witness for $(x, y_0, y_1, c_{sk})\in L^{\prime}$,
by running a $\Sigma_{OR}$-protocol (i.e., the OR-proofs of
$\Sigma$-protocols). 
The witness used by $P$ is $w$ such that $(x, w)\in
R_L$. We denote by $a_P, e_P, z_P$, the first-round, the
second-round, and the third-round message of the
$\Sigma_{OR}$-protocol of this stage, respectively.

\end{description}
\end{minipage}

\begin{minipage}[t]{6.4in} \small
\vspace{0.5cm}

\end{minipage}
\\
\hline
\end{tabular}
\caption{\label{withoutcw}
   $\Sigma_{OR}$-based  protocol variant without $c_{w}$}
\end{figure}
\end{center}

 Let $\hat{L}$ be
any $\mathcal{NP}$-language admitting a $\Sigma$-protocol that is
denoted by $\Sigma_{\hat{L}}$ (\emph{in particular, $\hat{L}$ can
be an empty set}). For an honest verifier $V$ with its public-key
$PK=(y_0, y_1)$, we define a new language $L=\{(\hat{x}, y_0,
y_1)| \exists w \ s.t. \ (\hat{x}, w)\in R_{\hat{L}}  \vee \exists
(w, b) \ s.t. \  y_b=f(w) \wedge b \in \{0, 1\}\}$. Note that for
any string $\hat{x}$ (whether $\hat{x}\in \hat{L}$ or not), the
statement ``$(\hat{x}, y_0, y_1) \in L$'' is always true as
$PK=(y_0, y_1)$ is honestly generated. Also note that $L$ is a
language that admits $\Sigma$-protocols (as $\Sigma_{OR}$-protocol
is itself a $\Sigma$-protocol). Now, we describe the concurrent
interleaving and  malleating attack,  in which $P^*$ successfully
convinces the honest verifier of the statement ``$(\hat{x}, y_0,
y_1) \in L$'' for \emph{any arbitrary} $poly(n)$-bit string
$\hat{x}$ (\emph{even when $\hat{x}\not \in \hat{L}$}) by
concurrently interacting with $V$ in two sessions as follows.

\begin{enumerate}
\item $P^*$ initiates the first session with $V$.  After receiving
the first-round message, denoted by  $a^{\prime}_V$, of the
$\Sigma_{OR}$-protocol  of Stage-1 of the first session on common
input $(y_0, y_1)$ (i.e.,  $V$'s public-key), $P^*$ suspends the
first session.

\item  $P^*$ initiates a second session   with $V$,  and works
just as the honest prover does in Stage-1 and Stage-2.  We denote
by $c_{sk}$ the Stage-2 message of the second session (i.e.,
$c_{sk}$ commits to $0^n$). When $P^*$ moves into Stage-3 of the
second session and needs to send $V$ the first-round message,
denoted by $a_P$, of the $\Sigma_{OR} $-protocol of Stage-3 of the
second session \emph{on common input $(\hat{x}, y_0, y_1,
c_{sk})$}, $P^*$ does the following:

\begin{itemize}

\item $P^*$ first runs the  SHVZK simulator of $\Sigma_{\hat{L}}$
(i.e.,  the $\Sigma$-protocol for $\hat{L}$) on $\hat{x}$ to get a
simulated conversation, denoted by $(a_{\hat{x}}, e_{\hat{x}},
z_{\hat{x}})$, for the (\emph{possibly false})  statement
``$\hat{x} \in \hat{L}$". Then,  $P^*$ runs the SHVZK simulator of
the underlying $\Sigma$-protocol for $\mathcal{NP}$ on $(y_0, y_1,
c_{sk})$ to get a simulated conversation, denoted by $(a_{sk},
e_{sk}, z_{sk})$, for the (false) statement ``$\exists (w, r_{sk},
b) \ s.t.\ c_{sk}=C(w, r_{sk}) \wedge y_b=f(w) \wedge b\in \{0,
1\}$".

\item $P^*$ sets $a_P=(a_{\hat{x}}, a^{\prime}_V, a_{sk})$ and
sends $a_P$ to $V$ as the first-round message of the
$\Sigma_{OR}$-protocol of Stage-3 of the second session, where
$a^{\prime}_V$ is the one received by $P^*$ in the first session.

\item After receiving the second-round message of Stage-3 of the
second session, denoted by $e_P$ (i.e., the random challenge from
$V$), $P^*$ sets $e^{\prime}_V=e_P\oplus e_{\hat{x}} \oplus
e_{sk}$
 and then suspends the second session.
\end{itemize}

\item $P$ continues the first session, and sends
$e^{\prime}_V=e_P\oplus e_{\hat{x}} \oplus e_{sk}$
 as the second-round message of the
$\Sigma_{OR}$-protocol of Stage-1 of the first session.

\item After receiving the third-round message of the
$\Sigma_{OR}$-protocol of Stage-1 of the first session, denoted by
$z^{\prime}_V$, $P^*$ suspends the first session again.

\item $P^*$ continues the execution of the  second session again,
and sends  $z_P=((e_{\hat{x}}, z_{\hat{x}}), (e^{\prime}_V,
z^{\prime}_V), (e_{sk}, z_{sk}))$ to $V$  as the last-round
message of the second session.

\end{enumerate}

 Note that  $(a_{\hat{x}}, e_{\hat{x}},
z_{\hat{x}})$ is an accepting conversation for the (possibly
false) statement ``$\hat{x}\in \hat{L}$",  $(a^{\prime}_V,
e^{\prime}_V, z^{\prime}_V)$ is an accepting conversation for
showing the knowledge of the preimage of either $y_0$ or $y_1$,
$(a_{sk}, e_{sk}, z_{sk})$ is an accepting conversation for the
statement ``$\exists (w, r_{sk}, b) \ s.t.\ c_{sk}=C(w, r_{sk})
\wedge y_b=f(w) \wedge b\in \{0, 1\}$", and furthermore
$e_{\hat{x}}\oplus e^{\prime}_V \oplus e_{sk}=e_P$.
 According to the description of $\Sigma_{OR}$ (presented in
Section  \ref{basic}), this means that, from the viewpoint of $V$,
$(a_P, e_P, z_P)$ is an accepting conversation  of Stage-3 of the
second-session on common input $(\hat{x}, y_0, y_1)$. That is,
$P^*$ successfully convinced $V$ of the statement ``$(\hat{x},
y_0, y_1) \in L$''  (\emph{even  for $\hat{x}\not \in \hat{L}$})
in the second session   \emph{but without knowing any
corresponding $\mathcal{NP}$-witness}.

\subsubsection{The attack against variant protocol without
$c_{sk}$}\label{attackwithoutcsk}

 The variant protocol without
$c_{sk}$ is re-depicted in Figure \ref{withoutcsk} (page
\pageref{withoutcsk}).

\begin{center}
 \begin{figure}[!t]
\begin{tabular} {|c|}
 \hline
\rule[-3mm]{0mm}{8mm}  \textbf{ $\Sigma_{OR}$-based  protocol
variant without $c_{sk}$ \quad $\langle P, V \rangle$}\\ \hline
\begin{minipage}[t]{6.4in} \small \textbf{Key Generation.}
Let $f: \{0, 1\}^n\rightarrow \{0, 1\}^n$ be any OWF, 
 where $n$ is the
security parameter.  Each verifier $V$ selects
 random strings   $s_0$, $s_1$ from $\{0, 1\}^n$, randomly selects a bit $b\leftarrow\{0, 1\}$,
 computes $y_b=f(s_b)$ and sets $y_{1-b}=f(s_{1-b})$. $V$  registers $PK=(y_0, y_1)$ in a public file $F$ as its public-key, and keeps $SK=s_b$
 as its secret-key.
    \end{minipage}\\ \hline

\begin{minipage}[t]{6.4in}
\begin{description}
\item [Common input.] An element $x\in L\cap\{0, 1\}^n$. Denote by
$R_L$ the corresponding $\mathcal{NP}$-relation for $L$. \item [P
private input.] An $\mathcal{NP}$-witness $w\in \{0, 1\}^n$ for
$x\in L$. Here, we assume w.l.o.g. that the witness for any $x\in
L\cap\{0, 1\}^n$ is of the same length $n$. 
\end{description}
\end{minipage}\\ \hline

\begin{minipage}[t]{6.4in}

\begin{description}
\item \item [Stage-1.] $V$ proves to $P$ that it knows the
preimage of either  $y_0$ or $y_1$, by running a 
$\Sigma_{OR}$-protocol on the input $(y_0, y_1)$ 
 in which $V$ plays the role of the knowledge prover.
The witness used by $V$ in this stage is $s_b$. Denote by $a_V,
e_V, z_V$, the first-round, the second-round and the third-round
message of the $\Sigma_{OR}$-protocol, respectively.

\item [Stage-2.] If $V$ successfully finishes Stage-1, $P$ does
the following: it computes $c_w=C(w, r_w)$, where $C$ is a
perfectly-binding commitment scheme and $r_w$ is  the randomness
used for commitments.

\item[Stage-3.]  Define a new $\mathcal{NP}$-language
$L^{\prime}=\{(x, y_0, y_1, c_w)| (\exists (w, r_w)\  s.t. \ c_w=
C(w, r_w) \wedge (x, w)\in R_L) \vee (\exists (w, b)\ s.t.\
y_b=f(w) \wedge b\in \{0, 1\})\}$. Then, $P$ proves to $V$ that it
knows a witness for $(x, y_0, y_1, c_w)\in L^{\prime}$, by running
a $\Sigma_{OR}$-protocol. 
 The witness used by $P$ is $(w, r_w)$. We denote by $a_P, e_P, z_P$, the first-round,
the second-round, and the third-round message of the
$\Sigma_{OR}$-protocol of this stage, respectively.

\end{description}
\end{minipage}

\begin{minipage}[t]{6.4in}
\vspace{0.5cm}

\end{minipage}
\\
\hline
\end{tabular}
\caption{\label{withoutcsk}   $\Sigma_{OR}$-based protocol variant
without $c_{sk}$}
\end{figure}
\end{center}

 Now,  we describe the concurrent interleaving and malleating
attack,  in which $P^*$ successfully convinces the honest verifier
of the statement ``$x\in L$",  for any $n$-bit string $x$ and
\emph{for any $\mathcal{NP}$-language $L$}, without knowing any
$\mathcal{NP}$-witness by concurrently interacting with $V$ in two
sessions as follows.

\begin{enumerate}
\item $P^*$ initiates the first session with $V$. After receiving
the first-round message, denoted by  $a^{\prime}_V$, of the
$\Sigma_{OR}$-protocol  of Stage-1 of the first session on common
input $(y_0, y_1)$ (i.e.,  $V$'s public-key), $P^*$ suspends the
first session.

\item  $P^*$ initiates a second session   with $V$,  and works
just as the honest prover does in Stage-1. In Stage-2 of the
second session, $P^*$ sends $c_w=C(0^n)$ (rather than $C(w)$ as
honest prover does). When $P^*$ moves into Stage-3 of the second
session and needs to send $V$ the first-round message, denoted by
$a_P$, of the $\Sigma_{OR}$-protocol of Stage-3 of the second
session \emph{on common input $(x,  y_0, y_1, c_w)$}, $P^*$ does
the following:

\begin{itemize}

\item $P^*$ first runs the  SHVZK simulator of the underlying
$\Sigma$-protocol for $\mathcal{NP}$ on common input $(x, c_w)$
 to get a simulated
conversation, denoted by $(a_{x}, e_{x}, z_{x})$, for the (false)
statement ``$\exists (w, r_w)\  s.t. \ c_w= C(w, r_w) \wedge (x,
w)\in R_L)$".

\item $P^*$ sets $a_P=(a_{x}, a^{\prime}_V)$ and sends $a_P$ to
$V$ as the first-round message of the $\Sigma_{OR}$-protocol of
Stage-3 of the second session, where $a^{\prime}_V$ is the one
received by $P^*$ in the first session.

\item After receiving the second-round message of Stage-3 of the
second session, denoted by $e_P$ (i.e., the random challenge from
$V$), $P^*$ sets $e^{\prime}_V=e_P\oplus e_x$ and then suspends
the second session.
\end{itemize}

\item $P$ continues the first session, and sends
$e^{\prime}_V=e_P\oplus e_x$
 as the second-round message of the $\Sigma_{OR}$-protocol of
Stage-1 of the first session.

\item After receiving the third-round message of the
$\Sigma_{OR}$-protocol of Stage-1 of the first session, denoted by
$z^{\prime}_V$, $P^*$ suspends the first session again.

\item $P^*$ continues the execution of the  second session again,
 and
sends  $z_P=((e_{x}, z_{x}), (e^{\prime}_V, z^{\prime}_V))$ to $V$
as the last-round message of the second session.

\end{enumerate}

 Note that  $(a_{x}, e_{x},
z_{x})$ is an accepting conversation for the (false) statement
``$\exists (w, r_w)\  s.t. \ c_w= C(w, r_w) \wedge (x, w)\in
R_L)$", $(a^{\prime}_V, e^{\prime}_V, z^{\prime}_V)$ is an
accepting conversation for showing the knowledge of the preimage
of either $y_0$ or $y_1$, and furthermore $e_{x}\oplus
e^{\prime}_V=e_P$. According to the description of $\Sigma_{OR}$
(presented in Section
 \ref{basic}), this means that, from the viewpoint of $V$,
$(a_P, e_P, z_P)$ is an accepting conversation  of Stage-3 of the
second-session on common input $x$. That is, $P^*$ successfully
convinced $V$ of the statement ``$x \in L$''
  \emph{but without knowing any
corresponding $\mathcal{NP}$-witness}.

\subsection{Practical instantiations}\label{apppractical}

In the (round-optimal)  practical instantiations of the efficient
CZK-CKE protocol, the verifier uses the sub-exponentially secure
DLP OWF in key-generation stage: $f_{p, q, g}(x)=g^x \mod p$,
where $p$ and $q$ are primes,  $p=2q+1$ and $|p|=n$,  and $g$ is
an element of $Z^*_p$ of order $q$. We also assume the (standard
polynomial-time) DDH assumption holds on the cyclic group indexed
by $(p, q, g)$ (i.e., the sub-group of order $q$ of $Z^*_p$). The
admissible common input is $x\in Z^*_p$ of order $q$ and the
corresponding witness is $w\in Z_q$ such that $g^w=x \mod p$. We
remark that 
 the parameters $(p, g, g)$, specifying the $f_{p, q, g}$ and the admissible common
inputs,  are  set outside the system.

The statistical  WIPOK of Stage-1 is replaced by the $\Sigma_{OR}$
of Schnorr's basic protocol for DLP \cite{S91}. The
perfectly-binding commitment scheme of Stage-2 is replaced by  the
DDH-based ElGamal (non-interactive) commitment scheme \cite{E85}
(recalled in  Section  \ref{basic}). To commit to a value $v\in
Z_q$, the committer randomly selects $u, r \in Z_q$, computes
$h=g^u \mod p$ and sends $(h, \bar{g}=g^r, \bar{h}=g^vh^r)$ as the
commitment.

For the practical $\Sigma$-protocol of Stage-3, by the
$\Sigma_{OR}$-technique we need the following two practical
$\Sigma$-protocols:

\begin{itemize}
\item A practical $\Sigma$-protocol that, given $x, c_w=(h,
\bar{g}, \bar{h})$, proves the knowledge of $(w, r)$ such that
$x=g^w \mod p$ and $\bar{g}=g^r \mod p$ and $\bar{h}=g^wh^r \mod
p$.

\item A practical $\Sigma$-protocol that, given $y_0, y_1,
c_{sk}=(h, \bar{g}_{sk}, \bar{h}_{sk})$, proves the knowledge $(w,
r)$ such that \textbf{either} $y_0=g^w
 \mod p$ and $\bar{g}_{sk}=g^r \mod p$ and $\bar{h}_{sk}=g^wh^r \mod
 p$ \textbf{or} $y_1=g^w
 \mod p$ and $\bar{g}_{sk}=g^r \mod p$ and $\bar{h}_{sk}=g^wh^r \mod
 p$.
\end{itemize}

Again, by the $\Sigma_{OR}$-technique, if we have a practical
$\Sigma$-protocol of the first type, then we can also have a
practical $\Sigma$-protocol of the second type. Thus, to get the
practical CZK-CKE implementation, all we need now is to develop a
practical $\Sigma$-protocol of the first type. Based on the
$\Sigma$-protocol for DLP \cite{S91}, such $\Sigma$-protocol is
described below.

\begin{description}
\item [Common input:]  $(p, q, g, x, h, \bar{g}, \bar{h})$, where
$x, h, \bar{g}, \bar{h}$ are all elements of order $q$ in $Z^*_p$.

\item [Prover's private input:] $w, r \in Z_q$ such that $x=g^w
\mod p$ and $\bar{g}=g^r \mod p$ and $\bar{h}=g^wh^r \mod p$.

\item [Round-1:] The prover $P$ randomly selects $t \in Z_q$,
computes $a_0=g^t \mod p$ and $a_1=h^t \mod p$, sends $(a_0, a_1)$
to the verifier $V$.

\item [Round-2:] $V$ responds back a random challenge $e$ taken
randomly from $Z_q$.

\item [Round-3:] $P$ computes $z_0=t+we \mod q$ and $z_1=t+re \mod
q$, and sends back $(z_0, z_1)$ to $V$.

\item [Verifier's decision:] $V$ accepts if and only if:
$g^{z_0}=a_0x^e \mod p$ and $g^{z_1}=a_0 \bar{g}^e \mod p$ and
$h^{z_1}=a_1 (\bar{h}/x)^e \mod p$.

\end{description}

We give a brief analysis of the above $\Sigma$-protocol:

\textbf{Special soundness}: From two accepting conversations
w.r.t. the \emph{same} Round-1 message,\\ $\{(a_0, a_1), e, (z_0,
z_1)\}$ and $\{(a_0, a_1), e^{\prime}, (z^{\prime}_0,
z^{\prime}_1)\}$, we can compute
$w=\frac{z_0-z^{\prime}_0}{e-e^{\prime}}$, and
$r=\frac{z_1-z^{\prime}_1}{e-e^{\prime}}$.

\textbf{Special HVZK:} The SHVZK simulator $S$ works as follows:
on a given random challenge $e\in Z_q$, it randomly selects $z_0,
z_1$ from $Z_q$, then it sets $a_0=g^{z_0}x^{-e}$ and
$a_1=g^{z_1}\bar{g}^{-e}=h^{z_1}(\bar{h}/x)^{-e}$.

We remark that, although the above  practical implementation is
for
specific number-theoretic language, 
 it is indeed very useful in practical scenarios.




\vspace{0.7cm}  \noindent \textbf{Acknowledgments.} We are
indebted to Frances F. Yao for numerous insightful  discussions
and suggestions on the early versions of this work. We thank
Giovanni Di Crescenzo, 
 Yehuda Lindell, 
Pino Persiano, Alon Rosen  and Ivan Visconti for  helpful
discussions.






\begin {thebibliography}{99}{



\bibitem{B01}
B. Barak.
\newblock{How to Go Beyond the Black-Box Simulation Barrier}.
\newblock In {\em {IEEE} Symposium on Foundations of Computer
Science}, pages 106-115, 2001.

\bibitem{BCNP04} B. Barak, R. Canetti, J. B. Nielsen and R. Pass.
\newblock{Universally Composable Protocols with Relaxed Set-Up
Assumptions}.
\newblock In {\em {IEEE} Symposium on Foundations of Computer Science}, pages 186-195,
2004.


\bibitem{BGGL01}
B. Barak, O. Goldreich, S. Goldwasser and Y. Lindell.
\newblock{Resettably-Sound Zero-Knowledge and Its Applications}.
\newblock In {\em {IEEE} Symposium on Foundations of Computer Science}, pages 116-125,
2001.

\bibitem{BPS06}
B. Barak, M. Prabhakaran, and A. Sahai.
\newblock{Concurrent Non-Malleable Zero-Knowledge}.
Cryptology ePrint Archive, Report No. 2006/355.  Extended abstract
appears in FOCS 2006.



 \bibitem{BG92}
 M. Bellare and O. Goldreich.
 \newblock{On Defining Proofs of Knowledge}
\newblock In {\em {E. F. Brickell (Ed.):  Advances in Cryptology-Proceedings of  CRYPTO 1992,
LNCS 740}}, pages 390-420,
  Springer-Verlag, 1992.

\bibitem{BG06}
M. Bellare and O. Goldreich.
\newblock{On Probabilistic versus Deterministic Provers in the Definition of Proofs Of
Knowledge}.
\newblock{Electronic Colloquium on Computational Complexity},
13(136), 2006. Available also from  Cryptology ePrint Archive,
Report No. 2006/359.

\bibitem{BS99}
M. Bellare and A. Sahai.
\newblock{Non-Malleable Encryption: Equivalence between Two Notions and an Indistinguishability-Based Characterization}.
\newblock In {\em {M. J. Wiener (Ed.):  Advances in Cryptology-Proceedings of  CRYPTO 1999, LNCS 1666}}, pages 519-536.
 Springer-Verlag, 1999. Full version appears in Cryptology ePrint
 Archive, Report No. 2006/228.








\bibitem{B82}
M. Blum.
\newblock{Coin Flipping by Telephone}.
\newblock In {\em proc. {IEEE} Spring COMPCOM}, pages 133-137, 1982.

\bibitem{B86}
M. Blum.
\newblock{How to Prove a Theorem so No One Else can Claim It}.
\newblock In {Proceedings of the International Congress of Mathematicians}, Berkeley, California, USA, 1986, pp. 1444-1451.








\bibitem{BCC88}
Brassard, D. Chaum and C. Crepeau.
\newblock{Minimum Disclosure Proofs of Knowledge}.
\newblock{\em Journal of Computer Systems and Science}, 37(2): 156-189, 1988.






\bibitem{CGGM00}
R. Canetti, O. Goldreich, S. Goldwasser and S. Micali.
\newblock{Resettable Zero-Knowledge}.
\newblock In {\em {ACM} Symposium on Theory of Computing},
  pages 235-244, 2000. Available from: http://www.wisdom.weizmann.ac.il/$\sim$oded/




\bibitem{CKPR02}
R. Canetti, J. Kilian, E. Petrank and A. Rosen.
\newblock{Black-Box Concurrent Zero-Knowledge Requires (Almost) Logarithmically Many Rounds}.
\newblock In {\em SIAM Journal on Computing}, 32(1): 1-47, 2002.


\bibitem{CLOS02}
R. Canetti, Y. Lindell, R. Ostrovsky and A. Sahai.
\newblock{Universally Composable  Two-Party and Multi-Party Secure
Computation}.
\newblock In {\em {ACM} Symposium on Theory of Computing},
  pages 494-503, 2002.

\bibitem{C96}
 R. Cramer.
 \newblock{Modular Design of Secure, yet Practical Cryptographic
 Protocols}, PhD Thesis, University of Amsterdam, 1996.




\bibitem{CDS94}
R. Cramer, I. Damgard and B. Schoenmakers.
\newblock{Proofs of Partial Knowledge and Simplified Design of Witness Hiding Protocols}.
\newblock In {\em {Y. Desmedt (Ed.):  Advances in Cryptology-Proceedings of  CRYPTO 1994, LNCS 893}}, pages 174-187.
  Springer-Verlag, 1994.



\bibitem{D00}
I. Damgard.
\newblock{Efficient Concurrent Zero-Knowledge in the Auxiliary String Model}.
\newblock In {\em {B. Preneel (Ed.):  Advances in Cryptology-Proceedings of  EUROCRYPT 2000, LNCS 1807}}, pages 418-430.
 Springer-Verlag, 2000.


\bibitem{D89}
I. Damgard.
\newblock{On the Existence of Bit Commitment Schemes and
Zero-Knowledge Proofs}.
\newblock In {\em {G. Brassard (Ed.):  Advances in Cryptology-Proceedings of  CRYPTO 1989, LNCS 435}}, pages 17-27.
Springer-Verlag, 1989.


\bibitem {D03}
 I. Damgard.
\newblock{Lecture Notes on Cryptographic Protocol Theory}.  BRICS,
Aarhus University, 2003.  Available from:
http://www.daimi.au.dk/$\sim$ivan/CPT.html





\bibitem{DPP93}
I. Damgard, T. Pedersen and B. Pfitzmann.
\newblock{On the Existence of Statistically-Hiding Bit Commitment
and Fail-Stop Signatures}. In Crypto 1993.

\bibitem{DL06}
Y. Deng and D. Lin.
\newblock{Resettable Zero Knowledge in the Bare Public-Key Model
under Standard Assumption}. Cryptology ePrint Archive, Report No.
2006/239.

\bibitem{DDL06}
Y. Deng, G. Di Crescenzo,  and D. Lin.
\newblock{Concurrently Non-Malleable Zero-Knowledge in the Authenticated Public-Key Model}.
Cryptology ePrint Archive, Report No. 2006/314, September 12, 2006.

\bibitem{DO99}
G. Di Crescenzo and R. Ostrovsky.
\newblock{On Concurrent Zero-Knowledge with Pre-Processing}.
\newblock In {\em {M. J. Wiener (Ed.):  Advances in Cryptology-Proceedings of  CRYPTO 1999, LNCS 1666}}, pages 485-502.
 Springer-Verlag, 1999.



 \bibitem{DPV04}
 G. Di Crescenzo, G. Persiano and I. Visconti.
 \newblock{Constant-Round Resettable Zero-Knowledge with
 Concurrent Soundness in the Bare Public-Key Model}.
 \newblock In {\em {M. Franklin  (Ed.):  Advances in Cryptology-Proceedings of  CRYPTO 2004, LNCS
3152}}, pages 237-253.
 Springer-Verlag, 2004.



\bibitem{DV05}
G. Di Crescenzo and I. Visconti.
\newblock{Concurrent Zero-Knowledge in the Public-Key Model}.
\newblock In {\em {L. Caires et al. (Ed.): ICALP 2005, LNCS 3580}},
pages 816-827. Springer-Verlag, 2005.



\bibitem{DVZ04}
G. Di Crescenzo, I. Visconti and Y. Zhao.
\newblock{Personal Communications, 2004}.

\bibitem{DDN00}
D. Dolev, C. Dwork and M. Naor.
\newblock{Non-Malleable Cryptography}.
\newblock{\em SIAM Journal on Computing}, 30(2): 391-437, 2000.
\newblock Preliminary version in {\em {ACM} Symposium on Theory of Computing},
  pages 542-552, 1991.



\bibitem{DNS98}
C. Dwork, M. Naor and A. Sahai.
\newblock{Concurrent Zero-Knowledge}.
\newblock In {\em {ACM} Symposium on Theory of Computing},
 pages 409-418, 1998.







\bibitem{DS98}
C. Dwork and A. Sahai.
\newblock{Concurrent Zero-Knowledge: Reducing the Need for Timing Constraints}.
\newblock In {\em {H. Krawczyk (Ed.):  Advances in Cryptology-Proceedings of  CRYPTO 1998, LNCS 1462}}, pages 442-457.
  Springer-Verlag, 1998.


 \bibitem{E85}
 T. El Gamal.
\newblock{A Public-Key Cryptosystem and Signature Scheme Based  on
Discrete Logarithms}.
\newblock{\em IEEE Transactions on Information Theory}, 31: 469-472, 1985.


\bibitem{F90}
U. Feige.
\newblock{Alternative Models for Zero-Knowledge Interactive
Proofs}. Ph.D Thesis, Weizmann Institute of Science, 1990.


\bibitem{FS89}
U. Feige and Shamir.
\newblock{Zero-Knowledge Proofs of Knowledge in Two Rounds}.
\newblock In {\em {G. Brassard (Ed.):  Advances in Cryptology-Proceedings of  CRYPTO 1989, LNCS 435}}, pages 526-544.
Springer-Verlag, 1989.




\bibitem{FFS88}U. Feige, A. Fiat and A. Shamir.
\newblock{Zero-knowledge Proof of Identity}.
\newblock {\em Journal of Cryptology},  1(2): 77-94, 1988.



\bibitem{GM00}
J. Garay and P. MacKenzie.
\newblock{Concurrent Oblivious Transfer}.
\newblock In {\em {IEEE} Symposium on Foundations of Computer Science}, pages 314-324,
2000.


\bibitem{G01}
O. Goldreich.
\newblock{\em Foundation of Cryptography-Basic Tools}. Cambridge University Press, 2001.


\bibitem{G02}
O. Goldreich.
\newblock{\em Foundations of Cryptography-Basic Applications}.
Cambridge University Press, 2002.





\bibitem{GK96} O. Goldreich and A. Kahan.
\newblock{How to Construct Constant-Round Zero-Knowledge  Proof Systems for $\mathcal{NP}$}.
\newblock{\em Journal of Cryptology}, 9(2): 167-189, 1996.

\bibitem{GKr96} O. Goldreich and H. Krawczyk.
\newblock{On the Composition of Zero-Knowledge Proof Systems}.
\newblock{\em SIMA Journal on Computing}, 25(1): 169-192, 1996.





\bibitem{GMR88}
S. Goldwasser, S. Micali and  R. L. Rivest.
\newblock{A Digital Signature Scheme Secure Against Adaptive Chosen
Message Attacks}.
\newblock {\em SIAM Journal on Computing}, 17(2): 281-308, 1988.

\bibitem{GMW86F}
O. Goldreich, S. Micali and A. Wigderson. \newblock{Proofs that
Yield Nothing but Their Validity and a Methodology of Cryptographic
Protocol Design}.
\newblock In {\em {IEEE} Symposium on Foundations of Computer Science}, pages 174-187,
1986.




\bibitem{GMW86C}
O. Goldreich, S. Micali and A. Wigderson.
\newblock{How to Prove all $\mathcal{NP}$-Statements in
Zero-Knowledge, and a Methodology of Cryptographic Protocol Design}.
\newblock In {\em {A. M. Odlyzko (Ed.):    Advances in Cryptology-Proceedings of  CRYPTO 1986, LNCS 263}}, pages 104-110,
Springer-Verlag, 1986.


\bibitem{GMW87}
O. Goldreich, S. Micali and A. Wigderson.
\newblock{How to Play any Mental Game-A Completeness Theorem for Protocols with Honest Majority}.
\newblock In {\em {ACM} Symposium on Theory of Computing},
 pages 218-229, 1987.


\bibitem{GMW91}
O. Goldreich, S. Micali and A. Wigderson.
\newblock{Proofs that Yield Nothing But Their Validity or All languages in $\mathcal{NP}$ Have Zero-Knowledge Proof Systems}.
\newblock {\em Journal of the Association for Computing Machinery}, 38(1): 691-729,
1991. Preliminary version appears in \cite{GMW86F,GMW86C}.


\bibitem{GMR85}
S. Goldwasser, S. Micali and  C. Rackoff.
\newblock{The Knowledge Complexity of Interactive Proof-Systems}
\newblock In {\em {ACM} Symposium on Theory of Computing},
  pages 291-304, 1985.







 \bibitem{GQ88}
L. Guillou and J. J. Quisquater.
\newblock{A Practical Zero-Knowledge Protocol Fitted to Security
Microprocessor Minimizing both Transmission and Memory}.
\newblock In {\em {C. G. G¨¹nther (Ed.):  Advances in Cryptology-Proceedings of  EUROCRYPT 1988, LNCS 330 }}, pages 123-128,
 Springer-Verlag, 1988.

 \bibitem {HILL99}
 J. Hastad, R. Impagliazzo, L. A. Levin and M. Luby.
 \newblock{Construction of a Pseudorandom Generator from Any One-Way
 Function}
\newblock {\em SIAM Journal on Computing}, 28(4): 1364-1396, 1999.

\bibitem{HR06}
I. Haitner and O. Reingold.
\newblock{Statistically-Hiding Commitment from Any One-Way
Function}. Cryptology ePrint Archive, Report No. 2006/436.

\bibitem{HHK05}
I. Haitner, O. Horvitz, J. Katz, C. Koo, R. Morselli and R.
Shaltiel.
\newblock{Reducing Complexity Assumptions for Statistically-Hiding
Commitments}. In Eurocrypt 2005.

\bibitem{HM96}
S. Halevi and S. Micali.
\newblock{Practical and Provably-Secure Commitment Schemes from
Collision-Free Hashing}. In Crypto 1996.

\bibitem{KLP05}
 Y. Kalai, Y. Lindell and M. Prabhakaran.
 \newblock{Concurrent Composition of Secure Protocols in the Timing Model}.
 \newblock In {\em {ACM} Symposium on Theory of Computing}, pages 644-653, 2005.

 \bibitem{K02}
 J. Katz.
 \newblock{Efficient Cryptographic Protocols Preventing
 ``Man-in-the-Middle" Attacks}. Ph.D Thesis, Columbia University,
 2002.

 \bibitem{K03}
 J. Katz.
 \newblock{Efficient and Non-Malleable Proofs of Plaintext Knowledge and Applications}.
\newblock In {\em {E. Biham (Ed.):  Advances in Cryptology-Proceedings of  EUROCRYPT 2003, LNCS 2656 }}, pages 211-228.
 Springer-Verlag, 2003.






\bibitem{K90}
J. Kilian.
\newblock{Uses of Randomness in Algorithms and Protocols}.
MIT Press, Cambridge, MA, 1990.




 \bibitem{L01}
Y. Lindell.
\newblock{Parallel Coin-Tossing and Constant-Round Secure Two-Party Computation}.
\newblock In {\em {J. Kilian (Ed.):  Advances in Cryptology-Proceedings of  CRYPTO 2001, LNCS 2139}}, pages 171-189.
  Springer-Verlag, 2001.


\bibitem{L03stoc}
Y. Lindell.
\newblock{Bounded-Concurrent Secure Two-Party
Computation Without Setup Assumptions}.
\newblock In {\em {ACM} Symposium on Theory of Computing},
 pages 683-692, 2003.



\bibitem{L04tcc}
Y. Lindell.
\newblock{Lower Bounds for Concurrent Self Composition}.
\newblock In {Theory of Cryptography (TCC) 2004,  LNCS 2951},
 pages 203-222, Springer-Verlag,  2004.

 \bibitem{Ljoc}
 Y. Lindell.
 \newblock{Lower Bounds and Impossibility Results for Concurrenet
 Self Composition}.
 \newblock{\em Journal of Cryptology}, to appear. Preliminary
 versions appear in \cite{L03stoc} and \cite{L04tcc}.


\bibitem{MP03}
D. Micciancio and  E. Petrank.
\newblock{Simulatable Commitments and Efficient Concurrent Zero-Knowledge}.
\newblock In {\em {E. Biham (Ed.):  Advances in Cryptology-Proceedings of  EUROCRYPT 2003, LNCS 2656 }}, pages 140-159.
 Springer-Verlag, 2003.

 \bibitem{MPR06}
 S.  Micali, R. pass and A. Rosen.
 \newblock{Input-Indistinguishable Computation}.
\newblock In {\em {IEEE} Symposium on Foundations of Computer Science}, pages
3136-145, 2006.


\bibitem{MR01a}
S. Micali and L. Reyzin.
\newblock {Soundness in the Public-Key Model}.
\newblock In {\em {J. Kilian (Ed.):  Advances in Cryptology-Proceedings of  CRYPTO 2001, LNCS 2139}}, pages 542--565.
  Springer-Verlag, 2001.






\bibitem{N91}
M. Naor.
\newblock{Bit Commitment Using Pseudorandomness}.
\newblock{\em Journal of Cryptology}, 4(2):
151-158, 1991.

\bibitem{NOVY98}
M. Naor, R. Ostrovsky, R. Venkatesan and  M. Yung.
\newblock{ Perfect Zero-Knowledge Arguments for NP Using Any One-Way Permutation}.
\newblock{\em Journal of Cryptology}, 11(2):
87-108, 1998.


\bibitem{NY90}
M. Naor and M. Yung.
\newblock{Public-Key Cryptosystems Provably Secure Against Chosen Ciphertext Attacks}.
\newblock In {\em {ACM} Symposium on Theory of Computing},
  pages 427-437, 1990.


   \bibitem{OPV06}
  R. Ostrovsky, G. Persiano and I. Visconti.
  \newblock{Concurrent Non-Malleable Witness Indistinguishability
  and Its Applications}.
\newblock{Electronic Colloquium on Computational Complexity},
13(95), 2006. Available also from  Cryptology ePrint Archive, Report
No. 2006/256.


\bibitem{OPV07} R. Ostrovsky, G. Persiano and I. Visconti.
\newblock{Constant-Round Concurrent NMWI and Its Relation to NMZK}.
Revised version of \cite{OPV06}, ECCC, March 2007.


\bibitem{P03C}
 R. Pass.
\newblock{On Deniabililty in the Common Reference String and Random
Oracle Models}.
\newblock In {\em {D. Boneh (Ed.):  Advances in Cryptology-Proceedings of  CRYPTO 2003, LNCS 2729}},
 pages 316-337, Springer-Verlag
 2003.

 \bibitem{PR05s}
 R. Pass and A. Rosen.
 \newblock{New and Improved Constructions of Non-Malleable
 Cryptographic Protocols}.
\newblock{\em SIAM Journal on Computing}, 38(2): 702-752 (2008).
\newblock  Preliminary version appears in  {\em {ACM} Symposium on Theory of Computing},
  pages 533-542, 2005.

\bibitem{PR05f}
R. Pass and A. Rosen.
\newblock{Concurrent Non-Malleable Commitments}.
\newblock{\em SIAM Journal on Computing}, 37(6): 1891-1925 (2008).
\newblock  Preliminary version appears in
\newblock In {\em {IEEE} Symposium on Foundations of Computer Science}, pages 563-572,
2005.




\bibitem{S91}
C. Schnorr.
\newblock{Efficient Signature Generation by Smart Cards}.
\newblock{\em Journal of Cryptology}, 4(3): 24, 1991.

\bibitem{V06}
I. Visconti. \newblock{Efficient Zero Knowledge on the Internet}.
\newblock{\em ICALP  2006,  LNCS 4052}, pages 22-33, Springer-Verlag.


\bibitem{Y86}
A. C. Yao.
\newblock{How to Generate and Exchange Secrets}.
\newblock In {\em {IEEE} Symposium on Foundations of Computer Science}, pages
  162-167, 1986.


\bibitem{YYZ07} A. Yao, F. Yao and Y. Zhao.
\newblock{Adaptive Concurrent Non-Malleability with Bare
Public-Keys}. Manuscript, 2007.

\bibitem{YYZ07eccc}
A. Yao, M. Yung and Y. Zhao.
\newblock{Concurrent Knowledge-Extraction in the Public-Key
Model}.
\newblock{Electronic Colloquium on Computational Complexity},
14(2), 2007.



\bibitem{YZ05}
M. Yung and Y. Zhao.
\newblock{Concurrently Knowledge-Extractable Resettable-ZK in the Bare Public-Key Model}.
\newblock{Electronic Colloquium on Computational Complexity},
12(48), 2005. Extended abstract appears \cite{YZ07}.

\bibitem{YZ07}
 M. Yung  and Y.   Zhao .
 \newblock{Generic and practical resettable zero-knowledge in the bare public-key model}.
\newblock In {\em {M. Naor (Ed.):  Advances in Cryptology-Proceedings of  EUROCRYPT 2007, LNCS 4515}},
 pages 116-134, Springer-Verlag,  2007.



\bibitem{YZ06}
M. Yung and Y. Zhao.
\newblock{Interactive Zero-Knowledge with Restricted Random
Oracles}.
\newblock In {\em {S. Halevi and T. Rabin (Ed.):  Theory of Cryptography (TCC) 2006,  LNCS 3876}},
 pages 21-40, Springer-Verlag,  2006.




\bibitem{Z03}
  \newblock{Concurrent/Resettable Zero-Knowledge With Concurrent Soundness in  the Bare Public-Key
   Model and Its Applications.} Unpublished manuscript, appears in
   Cryptology ePrint Archive,  Report  2003/265.






}
 \end {thebibliography}
\end{document}